1

# Control of an Unmanned Surface Vehicle with Uncertain Displacement and Drag

Wilhelm B. Klinger, Ivan R. Bertaska, Karl D. von Ellenrieder and M. R. Dhanak

*Abstract*— **Experimental testing of an unmanned surface vehicle (USV) has been performed to evaluate the performance of two low-level controllers when displacement and drag properties are time-varying and uncertain. The USV is a 4.3 meter long, 150 kilogram wave adaptive modular vessel (WAM-V) with an inflatable twin hull configuration and waterjet propulsion. Open loop maneuvering tests were conducted to characterize the dynamics of the vehicle. The hydrodynamic coefficients of the vehicle were determined through system identification of the maneuvering data and were used for simulations during control system development. The resulting controllers were experimentally field tested on-water. Variable mass and drag tests show that the vehicle is best controlled by a model reference adaptive backstepping speed and heading controller. The backstepping controller developed by Liao et. al (2010) is modified to account for an overprediction of necessary control action and motor saturation. It is shown that when an adaptive algorithm is implemented for the surge control subsystem of the modified backstepping controller, the effects of variable mass and drag are mitigated.**

## I. INTRODUCTION

A N understanding of a vehicle's maneuverability and dynamic performance characteristics is important for the development of an automatic control system that will command its motion and can also serve as a guide for the future design of similar vehicles. A potential application of unmanned surface vehicles (USV's) is the automatic launch and recovery (ALR) of an autonomous underwater vehicle (AUV) [27], [29], [33], [34]. A good closed loop controller for a USV used in such operations needs to allow for the changes in displacement and vehicle resistance associated with launch/recovery of the AUV payload. While gain-scheduling or other methods may be used when an AUV is either completely onboard the USV, or after it has been launched, the transition time during launch/recovery requires special considerations in developing automatic controllers because the magnitude and direction of the drag force due to the AUV and any associated docking/recovery system may be unpredictable.

Several alternatives for speed and heading control of a small twin-hulled unmanned surface vehicle with time-varying and uncertain displacement and drag are presented. Experimental on-water test results are presented for a USV under the variable and

Manuscript received Month day, 2015. This work was supported by the U.S. Office of Naval Research under Grant N000141110926, managed by K. Cooper. I. R. Bertaska gratefully acknowledges the support of the Link Foundation Ocean Engineering and Instrumentation Ph.D. Fellowship Program. W. Klinger, I. Bertaska, K. D. von Ellenrieder and M. R. Dhanak are with the Department of Ocean & Mechanical Engineering, Florida Atlantic University, Dania Beach, FL 33004-3023 USA (phone:+1-954-924-7232; fax:+1-954-924-7270; e-mail: ellenrie@fau.edu).

uncertain displacement and drag conditions expected during an ALR. The main scholarly outcomes of this work are the development and experimental validation of low level automatic control systems for a USV during an ALR as it transitions from carrying to towing an AUV, and also after deploying the AUV and operating independently.

The paper is organized as follows. Background for ALR and adaptive control for USVs is presented in Section II. The particulars of the vehicle, including its design and instrumentation, are described in Section III. In Section IV, we characterize the maneuvering performance of the vehicle. We develop controls oriented state space models of the vehicle through the system identification of experimental data in Section V. Three control alternatives for speed and heading control are explored in Section VI. In Section VII, we compare the controller's performance for time-varying and uncertain displacement and drag conditions. Finally, in Section VIII, some concluding remarks are given regarding the implications of our results.

## II. RELATED WORK

There have been a substantial number of studies on the nonlinear control of ships, boats, and unmanned surface vehicles, including [5],[39] ,[15],[10],[21],[17],[24]. However, as pointed out in [5] and [8], there is a lack of comprehensive and practical control laws that are robust to the uncertainties in vessel dynamics and large disturbances associated with actual on-water conditions. Additionally, as elaborated upon in [5], few studies attempt experimental implementation on real USV platforms.

Recent work in controls for unmanned surface vehicles incorporate nonlinearities within the equations of motion to produce nonlinear control algorithms that are able to achieve a wider performance envelope than their linear counterparts [4]. These include tracking and setpoint control utilizing sliding mode methods, as in [6] and [25], as well as integrator backstepping as proposed in [19]. Although various algorithms have been put forward, few have had the opportunity to be validated on full-size USVs in the presence of environmental disturbances and equipped with traditional navigation sensors, rather than lab-based analogs (e.g. overhead cameras in lieu of compass/GPS/IMU sensor suite) [9]. Some exceptions include [39],[8],[7],[9],[35],[31], [32], and [3]. In [39], the authors used the backstepping method using feedback linearization and Lyapunov's direct method to create a speed-scheduled algorithm that was field-tested on a USV in riverine operations. In [3], a sliding mode controller was developed for an amphibious surface vehicle navigating the transitional area between the shore and open ocean, where breaking waves cause large nonlinearities that may not be adequately modeled *in situ*. Automatically generated behaviors for USVs were presented in [8], [7], and [40] where three setpoint controllers were proposed and validated in the field. A controller from this paper is incorporated here in Section VI.

Adaptive control is also an active area of research for unmanned marine vehicles. Skjetne et al. used adaptive control to determine the added mass hydrodynamic coefficients while performing free-running maneuvering experiments in [36] to create a nonlinear maneuvering model suited to either tracking or dynamic positioning control. The methods behind backstepping and adaptive control were combined in [23], using a Nomoto-like steering model that neglected the coupling of yaw rate with sway velocity commonly found in USVs. A recent method of handling uncertain displacement and drag terms is presented in [4], where model predictive control is used to stabilize the vehicle's course under large mass variations upwards of 50%. It should be noted that the authors used a linearized model for plant dynamics, which may not fully capture the dynamics of the vehicle at high speeds. Aguiar and Hespanha proposed an approach utilizing supervisory switching control to handle parametric uncertainties in an underactuated marine vehicle [1]. Although favorable simulation results were presented for both an underactuated hovercraft and an AUV, the use of finite number of system models could potentially lead to large errors if parameters' range is infinite. Aguiar and Pascoal extended this work in parametric uncertainty in [2], where an adaptive backstepping controller was derived to estimate poorly modeled hydrodynamic parameters in an AUV with an unconventional hull form. Kragelund et al [22] proposed three different adaptive controllers – a gain scheduled PID controller, a model reference adaptive controller and an $L_1$ adaptive controller – tested on a SeaFox USV to mitigate the effects of the transitional region between displacement and planing modes of operations.

The work presented here extends the results of previous studies to include the effects of the adaptive control on course-keeping relating to vehicle bearing, as well as using a nonlinear controller derived from a nonlinear maneuvering model.

## III.   THE WAM-V USV14

### A. Overview

The WAM-V USV14 (Fig. 1) is a twin hull pontoon style vessel designed and built by Marine Advanced Research, Inc. of Berkeley, CA USA. The vessel structure consists of two inflatable pontoons, two motor pods, a payload tray, and two supporting arches. The WAM-V class of vehicles is designed to provide a stable platform that can mitigate the effects of waves [11], [16],[30]. The vehicle's physical characteristics are shown in TABLE I. The WAM-V demonstrates appreciable motion in surge, sway, and yaw during operation, but can only apply control force and torque in surge and yaw. Because the steering scheme can only actuate in two of the three appreciable degrees of freedom, the USV is underactuated.

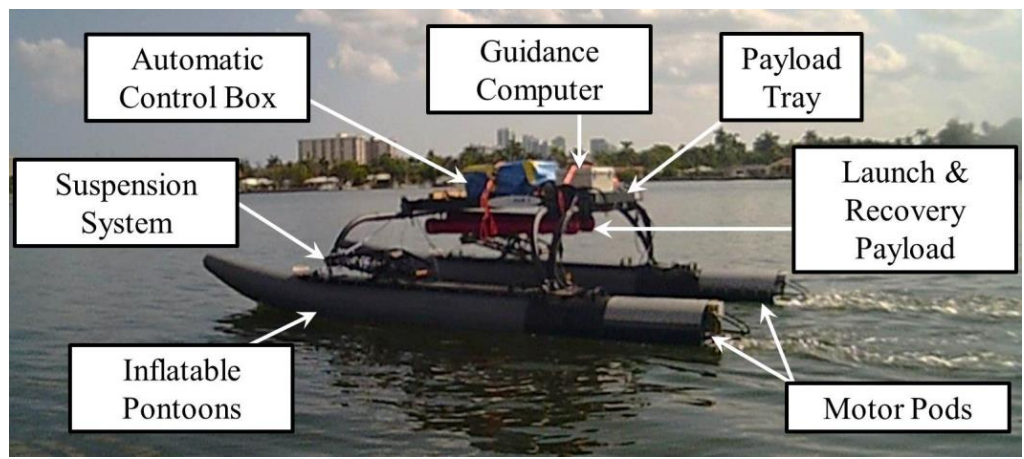

Fig. 1: The WAM-V USV14 Unmanned Surface Vehicle underway in the Intracoastal Waterway (ICW).

TABLE I

PRINCIPLE CHARACTERISTICS OF THE WAM-V14 USV. THE LOCATION OF THE "KEEL" IS TAKEN AS
THE BOTTOM OF THE PONTOONS. W.R.T. IS AN ACRONYM FOR THE PHRASE "WITH RESPECT TO".

| Parameter | Value |
|---|---|
| Length Overall (L) | 4.29 [m] |
| Length on the Waterline (LWL) | 3.21 [m] |
| Draft (aft and mid-length)(lightship) | 0.127 and 0.105 [m] |
| Beam Overall (BOA) | 2.20 [m] |
| Beam on the Waterline (BWL) | 2.19 [m] |
| Depth (keel to pontoon skid top) | 0.39 [m] |
| Area of the Waterplane (AWP) | 1.1 [m$^2$] |
| Centerline-to-centerline Side Hull Separation (B) | 1.83 [m] |
| Length to Beam Ratio (L/B) | 2.34 |
| Volumetric Displacement ($\nabla$) (lightship) | 0.34 [m$^3$] |
| Mass ($m$) | 150 [kg] |
| Longitudinal Center of Gravity (LCG) w.r.t. aft plane of engine pods | 1.27 [m] |

*B. Waterjet Propulsion System*

The waterjet configuration is similar to that found on the stock version of the WAM-V USV12 [41], except that each of the USV14 motor pods contains an electric-powered motor, rather than a gasoline-powered engine. The motor pods attach aft of the inflatable pontoons with a hinged connection at the bottom and a flexible O-ring connection at the top, which allows rotation about the bottom connection. The design is meant to ensure the motor pods' continuous alignment with the water, despite vehicle pitching motion, which maintains flow through the waterjets. Each motor is powered by eight lithium polymer batteries, wired in parallel, that provide a nominal voltage of 22.2 V. The motors are NeuMotors 2230, and are rated for a maximum of 22,000 RPM. The nozzles of the waterjet have been fixed so that they point directly aft, and the vehicle is steered using differential thrust. The waterjet reversing buckets are not used, such that thrust is only developed in the forward direction.

*C. Guidance, Navigation, and Control Electronics*

In order to conduct development and testing of autonomous control for the WAM-V USV14, a guidance, navigation and control (GNC) system is implemented. The GNC system is housed in a plastic, water-resistant box and contains a motherboard, single-board computer, inertial measurement unit (IMU) with global positioning system (GPS) capability, tilt-compensated digital compass, radio frequency (RF) transceiver, and pulse width modulation (PWM) signal generator. A detailed summary of the GNC system can be found in [7] and [26]. For the purpose of this project, the key components of the GNC system are the sensor suite (IMU/GPS and digital compass), single-board computer, and RF transceiver. The IMU/GPS is an XSENS MTi-G sensor , which is used to measure the position and orientation of the USV during operations. This information is relayed to the single board computer, where the low level control code is implemented, and maneuvering data are logged. The GPS is specified to provide up to 2.0 meter accuracy, depending on weather and satellite availability. The digital compass is used to record vehicle heading and has a resolution of 0.1 degrees. The RC receiver allows the user the option of maneuvering the USV under remote manual control or by autonomous navigation. A hand-held remote control is used both for operating the thrusters and reversing buckets and also for switching between RC and autonomous modes.

## IV. MANEUVERING CHARACTERISTICS

Maneuvering tests were performed in sheltered, calm water sections of the US Intracoastal Waterway (ICW) in Dania Beach and Hollywood, FL USA. The vehicle experienced only minor wind-generated waves. Wind speed and direction were collected during all testing. Wind-generated forces and moments do not appear to have had any effect on the test results.

The total weight (displacement) of the system will affect both the inertial properties and, through changes in hull submergence (draft), the hydrodynamic resistance of the vehicle. Thus, the maneuvering characteristics are dependent on how heavily laden the vehicle is. Here, three different loading conditions (total vehicle displacements), which have been termed "slick", "lightship" and "full displacement", were tested:

- **Slick:** The vehicle displacement with only GNC hardware onboard – the minimum loading to run autonomously. This configuration includes neither the simulated ALR retrieval mechanism, nor the AUV mock-up (see Section VII.A). The total displacement is approximately 150 kg.

- **Lightship:** The vehicle displacement with both the GNC hardware and the simulated ALR retrieval mechanism aboard, but not the AUV mock-up. This arrangement would correspond to the loading condition of the USV after a launch and before a recovery of the AUV. The ALR system has a mass of approximately 70 kg, so that the total displacement of the vehicle is about 220 kg.

- **Full Displacement:** The loading condition with the GNC hardware, simulated ALR retrieval mechanism, and AUV mock-up aboard. This corresponds to the weight of the USV before and after the ALR procedure. To simulate the weight of the AUV, an additional mass of 26 kg is added to the lightship system, such that its total displacement is about 246 kg.

### A. Acceleration Tests

Acceleration tests were conducted to obtain the steady state speeds associated with the available range of motor commands. The vehicle started in a stationary position, with the surge and sway velocities, and the yaw rate as close to zero (as possible on the water), and was accelerated to a specified throttle command on both motors for 60 seconds. The test was conducted for a throttle range of 60-100%. Preliminary experimental results for the slick acceleration test are presented in [20]. Fig. 2 shows the results of two full throttle acceleration tests for the lightship (no AUV in payload) and full displacement (AUV in payload) vehicle conditions. It can be seen from the data in Fig. 2 that the maximum surge velocity of the vessel decreases as the payload is increased. The average maximum surge speed for the lightship and full displacement cases is 2.8 m/s and 2.5 m/s, respectively.

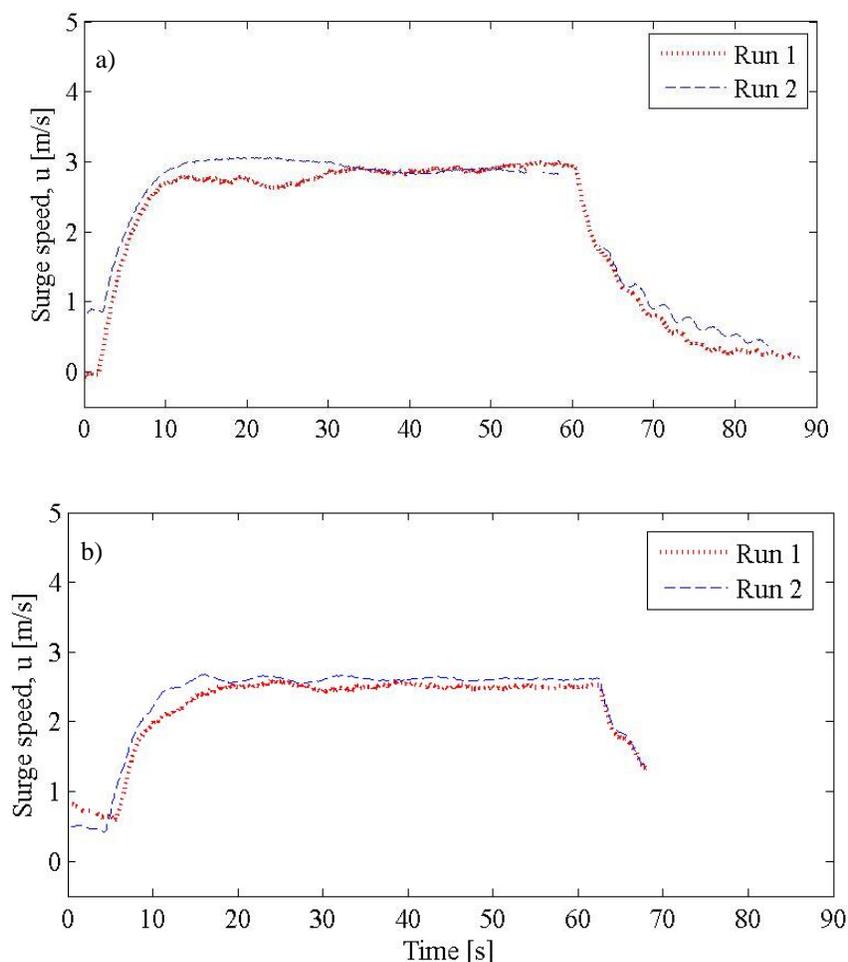

Fig. 2: Vehicle surge velocity for two full throttle acceleration tests using the automatic control box with preprogrammed motor commands. Vehicle conditions were: a) Lightship and b) Full Displacement.

The results of the acceleration tests provided a means to approximate a forward drag model for each displacement condition of the vessel. In order to relate motor commands to drag, a bollard-pull test was conducted. The test consisted of tethering the WAM-V USV14 to a fixed object (a pier) with a scale in line with the tether. The speed of the motors was slowly increased to full throttle; the maximum force measured is $T_{wj\_max} = 204$ Newtons. Repetitions of this test at multiple throttle commands indicated that thrust scales linearly with motor command when the forward speed of the vessel is zero. The results of the bollard-pull test were then incorporated into a linear model for thrust output as a function of input motor command. As the vehicle was tethered during these tests, the effect of surge speed on thrust production is ignored in this model:

$$T_{wj} = \frac{n}{n_{max}} T_{wj\_max}. \tag{1}$$

Here $n$ is the commanded motor speed in Hz; $n/n_{max}$ can be interpreted as a commanded percentage of the maximum thrust. Using this linear model, experiments were conducted at steady state forward speeds $u$, where it was assumed that the USV was operating at the "thrust equals drag" condition. With a drag model composed of both quadratic and linear terms, the thrust equals drag $D$ condition would give:

$$T_{\text{wj\_max}} \frac{n}{n_{\max}} = X_{u|u|} u|u| + X_u u. \tag{2}$$

$X_{u|u|}$ and $X_u$ were then determined from curvefits to the experimental data in Fig. 3.

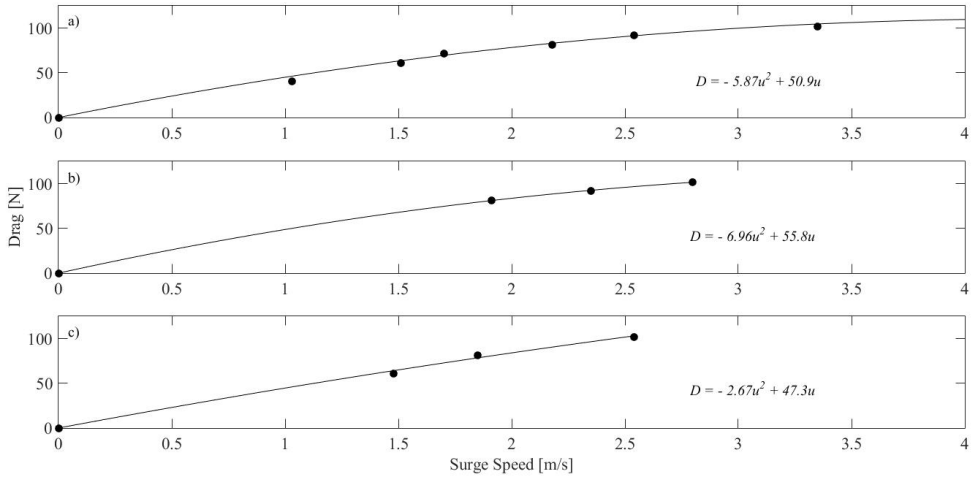

Fig. 3: Surge direction drag model for the WAM-V USV14 in three displacement conditions: a) slick, b) lightship, and c) full-displacement. The points represent steady state values of thrust equals drag ($D$) for a given speed ($u$).

Because (1) neglects the effect of surge speed on thrust developed, the drag model does not capture the characteristic variation of drag with surge speed that might be expected, e.g. $D \sim u^2$. However, use of the $X_{u|u|}$ and $X_u$ terms found from these experiments should intrinsically reproduce the surge speed effects when the desired thrust is determined using (1).

### 1) Alternative Thruster Models

A second possible thruster model, which has been developed as an analog to conventional propeller models, was also explored. In conventional propellers, the curves of thrust coefficient

$$K_T \equiv \frac{T_p}{\rho n^2 d^4},$$

versus advance ratio

$$J \equiv \frac{u}{nd},$$

are approximately linear such that

$$K_T(J) = -\alpha_1 J + \alpha_2.$$

Here, $T_p$ is propeller thrust, $d$ is propeller diameter, $\rho$ is water density and $\alpha_1, \alpha_2 > 0$ are constants. Using this model the thrust can be rewritten as

$$T_p = -\alpha_1 \frac{u}{nd} \rho n^2 d^4 + \alpha_2 \rho n^2 d^4,$$

or

$$T_p = a_1(\rho d^3)un + a_2(\rho d^4)n^2.$$

Here the thrust developed varies quadratically with the rotational rate of the propeller and linearly with the surge speed of the vessel. The authors have not come across similar waterjet models in the literature, but it seems that such a model might be appropriate. Examining a representative plot of waterjet pump head coefficient $C_H$ vs. capacity coefficient $C_Q$ (Fig. 4), one can see that an almost linear, relation exists between the pressure head coefficient

$$C_H \equiv \frac{gH}{(nd)^2},$$

(where $g$ is gravitational acceleration, $H$ is the pressure head across the waterjet, and $d$ is the diameter of the waterjet impeller) and the capacity coefficient

$$C_Q \equiv \frac{Q_f}{nd^3},$$

(where $Q_f$ is the volumetric flow rate through the waterjet). Note that $C_H$ and $C_Q$ are often combined with measured waterjet pump data to estimate the steady state waterjet impeller shaft speed (see [12], for example).

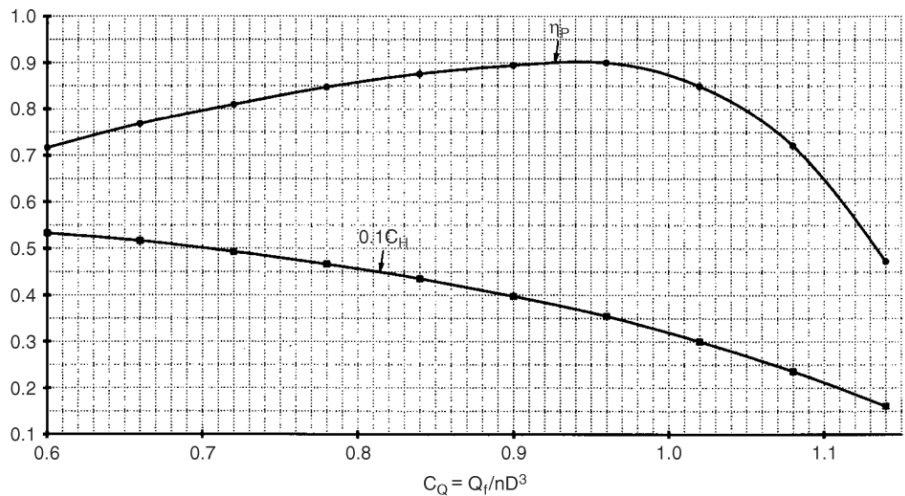

Fig. 4: Example pump diagram showing pump efficiency $\eta_P$ and head coefficient $C_H \equiv gH/(nd)^2$ as a function of the capacity coefficient $C_Q \equiv Q_f/nd^3$. $n$=shaft revolutions per second, $d$ = impeller diameter (reproduced from [13]).

Following a similar development as used above for a conventional propeller, we have

$$C_H(C_Q) = -\alpha_1 C_Q + \alpha_2,$$

which can be rewritten as

$$\rho g H = -\alpha_1 \frac{\rho Q_f}{nd^3}(nd)^2 + \alpha_2 \rho(nd)^2.$$

Scaling the volumetric flow rate through the waterjet as $ud^2$, and assuming that the thrust generated is proportional to the pressure head across the waterjet multiplied by the square of the characteristic diameter of the waterjet we get

                                                                9

$$T_{wj} \sim \rho d^2 \left[ -\alpha_1 \frac{ud^2}{nd^3}(nd)^2 + \alpha_2 (nd)^2 \right],$$

which has the same form as the thrust developed on a propeller

$$T_{wj} \sim a_1(\rho d^3)un + a_2(\rho d^4)n^2.$$

Curvefits of this relation to the data collected during our field tests indicate that the model works well for the lightship condition, but does not match as well for the slick or full displacement conditions. This waterjet thrust model may be appropriate in certain circumstances and appears to be worthy of additional investigation.

### B. Maneuvering Tests

A modified on-water zig-zag test was conducted to obtain motion data. The traditional zig-zag test specifies rudder commands for a certain heading range [18], but for a vehicle steered using differential thrust, the test must be conducted by varying the motor speeds. Motor commands were alternated between 0 and 100% power, with the motors $180^o$ out of phase. Each motor setting was held for 7 seconds. Trajectory and yaw rate ($r$) results for three tests in the lightship condition are shown in Fig. 5. Yaw rate is of interest when conducting the zig-zag test because it is later used in model development, specifically in determining the maneuvering coefficients pertaining to how fast the vessel changes heading. This concept is explored further in Section V. Several runs of each maneuvering test were carried out to show a convergence of the results and to inspire confidence in the results before using them to develop the dynamic model of the vehicle.

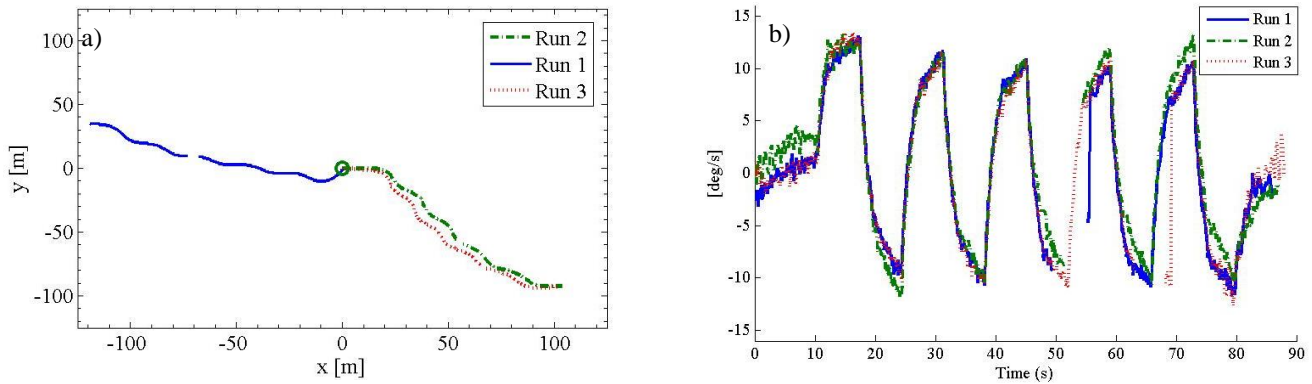

Fig. 5. a) Trajectory and b) yaw rate time histories for three zig-zag tests conducted in the lightship displacement condition. The maximum yaw rate was 13 deg/s. In a) the starting position is marked by the circle at the origin.

## V. SYSTEM IDENTIFICATION

### A. Equations of Motion

Data collected from the acceleration and zig-zag tests in Sections IV.A and IV.B, respectively, were used to develop three degree of freedom state space maneuvering models for the vehicle in full displacement and lightship displacement conditions. Following the notation developed in [13] and [14], a three degree of freedom (surge, sway, and yaw) vehicle model is used, as presented in

[20] and [21], where

$$\boldsymbol{M}\dot{\boldsymbol{v}} + \boldsymbol{C}(\boldsymbol{v})\boldsymbol{v} + \boldsymbol{D}(\boldsymbol{v})\boldsymbol{v} = \boldsymbol{\tau} \tag{3}$$

$$\dot{\boldsymbol{\eta}} = \begin{bmatrix} \dot{x} & \dot{y} & \dot{\psi} \end{bmatrix}^T, \tag{4}$$

and

$$\boldsymbol{v} = [u \; v \; r]^T. \tag{5}$$

Here, the $\dot{\boldsymbol{\eta}}$ describes the vehicle's North ($\dot{x}$) East ($\dot{y}$) linear velocities and the Z-axis angular velocity ($\dot{\psi}$) in an inertial reference frame ($x, y, \psi$). $\boldsymbol{v}$ contains the vehicle surge velocity ($u$), sway velocity ($v$) and yaw rate ($r$) in the body fixed frame. Fig. 6 illustrates the two coordinate systems and the motor forces acting on the vehicle.

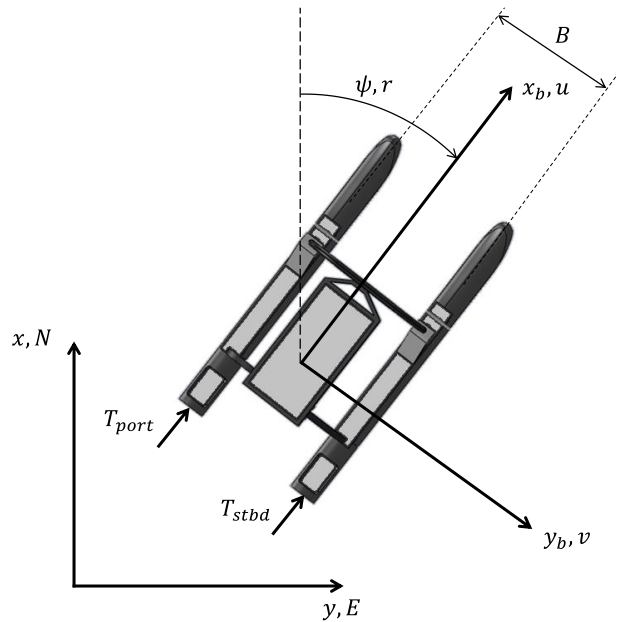

Fig. 6. Top view of WAM-V USV14 with body-fixed coordinate system overlaid. $x_b$ and $\boldsymbol{y_b}$ denote vessel surge and sway axes, respectively. $T_{port}$ and $T_{stbd}$ denote the port and starboard waterjet thrust.

The transformation matrix for converting from body-fixed to earth-fixed frames is given by:

$$\boldsymbol{J}(\boldsymbol{\eta}) = \begin{bmatrix} \cos\psi & -\sin\psi & 0 \\ \sin\psi & \cos\psi & 0 \\ 0 & 0 & 1 \end{bmatrix}. \tag{6}$$

$\boldsymbol{M}$ is an inertia tensor that is the sum of a rigid body mass matrix, $\boldsymbol{M_{RB}}$, and an added mass matrix, $\boldsymbol{M_A}$:

$$\boldsymbol{M} = \boldsymbol{M_{RB}} + \boldsymbol{M_A} = \begin{bmatrix} m - X_{\dot{u}} & 0 & -my_G \\ 0 & m - Y_{\dot{v}} & mx_G - Y_{\dot{r}} \\ -my_G & mx_G - N_{\dot{v}} & I_z - N_{\dot{r}} \end{bmatrix}, \tag{7}$$

where, $m$ denotes the mass of the USV-AUV system, $x_G$ and $y_G$ represent the coordinates of the vessel center of mass in the body-fixed frame, and $I_z$ denotes moment of inertia about the $z_b$-axis. Several of the terms in the mass matrix utilize SNAME (1950) [38] nomenclature for representing a force or moment created by motion in a specific degree of freedom, where $X$ and $Y$ correspond to force in the $x_b$ and $y_b$ directions, respectively, and $N$ denotes a moment about the $z_b$-axis (vertical axis). The subscript on each coefficient denotes the cause of the force/moment. It is pertinent to note that the added mass in the surge direction is modeled as 16% of the total mass of the vehicle, whereas theoretically the value should be less than 5%. This change is made so that the simulation outputs a forward acceleration curve that matches the experimental data.

$C(v)$ is a Coriolis matrix, which includes the sum of a rigid body ($C_{RB}$) and added mass ($C_A$) matrix:

$$C(v) = C_{RB} + C_A = \begin{bmatrix} 0 & 0 & -m(x_G r + v) \\ 0 & 0 & -m(y_G r - u) \\ m(x_G r + v) & m(y_G r - u) & 0 \end{bmatrix}$$
$$+ \begin{bmatrix} 0 & 0 & \dfrac{Y_{\dot{v}}v + \left(\frac{Y_{\dot{r}}+N_{\dot{v}}}{2}\right)r}{200} \\ 0 & 0 & -X_{\dot{u}}u \\ \dfrac{-Y_{\dot{v}}v - \left(\frac{Y_{\dot{r}}+N_{\dot{v}}}{2}\right)r}{200} & X_{\dot{u}}u & 0 \end{bmatrix} \quad (8)$$

The model designates the body-fixed origin at the center of gravity and assumes horizontal symmetry, making $x_G = 0$ and $y_G = 0$. The factor of 1/200 in the upper-rightmost and lower-leftmost matrix positions of $C_A$ was applied during model development to better fit the experimental data and does not appear in [13] and [14]. This scale factor was necessary to fix an issue with the simulation where the USV could not pull out of a turn after one had been initiated, even with maximum torque applied. The original scale factor was 2 to represent each demihull, and was manually incremented until the turning behavior of the simulation model was deemed acceptable. $D(v)$ is the summation of linear and nonlinear drag matrices:

$$D(v) = D_l + D_n \quad (9)$$

where,

$$D_l = -\begin{bmatrix} X_u & 0 & 0 \\ 0 & Y_v & Y_r \\ 0 & N_v & N_r \end{bmatrix} \quad (10)$$

and,

$$D_n = -\begin{bmatrix} X_{u|u|}|u| & 0 & 0 \\ 0 & Y_{v|v|}|v| + Y_{v|r|}|r| & Y_{r|v|}|v| + Y_{r|r|}|r| \\ 0 & N_{v|v|}|v| + N_{v|r|}|r| & N_{r|v|}|v| + N_{r|r|}|r| \end{bmatrix}. \quad (11)$$

The surge direction drag terms $D_{port}$ and $D_{stbd}$ are modeled using a polynomial curve fit derived from experimental testing. A coordinate transformation is carried out to obtain the velocities of each individual pontoon hull because they are offset from the CG. These transformed velocities are used in the drag model below as $u_{port}$ and $u_{stbd}$:

$$D_{port} = \left(\frac{X_{u|u|}}{2}\right)|u_{port}|u_{port} + \left(\frac{X_u}{2}\right)u_{port} \quad (12)$$

$$D_{stbd} = \left(\frac{X_{u|u|}}{2}\right)|u_{stbd}|u_{stbd} + \left(\frac{X_u}{2}\right)u_{stbd} \tag{13}$$

The values of $X_{u|u|}$ and $X_u$ vary based on the displacement condition of the vehicle and are shown in TABLE II. Incorporating the moment created by the two drag forces $D_{port}$ and $D_{stbd}$, the term

$$\left(D_{stbd} - D_{port}\right)B/2 \tag{14}$$

is added to the yaw moment (not modeled in $\boldsymbol{D}$).

Finally, $\boldsymbol{\tau}$ is a vector of the forces and moments generated by the port and starboard waterjets:

$$\boldsymbol{\tau} = \begin{pmatrix} \tau_x \\ \tau_y \\ \tau_z \end{pmatrix} = \begin{bmatrix} (T_{port} + T_{stbd}) \\ 0 \\ (T_{port} - T_{stbd})B/2 \end{bmatrix}. \tag{15}$$

As shown in Fig. 6, $T_{port}$ and $T_{stbd}$ denote the thrust from the port and starboard waterjets, respectively, and $B/2$ is the transverse distance from the centerline of the vessel to the centerline of each hull. The forces caused by wind and waves are neglected.

TABLE II

SURGE DRAG COEFFICIENTS FOR THE WAM-V USV14.

| Displacement Condition | $X_{u|u|}$ | $X_u$ |
|---|---|---|
| Slick | -5.8722 | 50.897 |
| Lightship | -6.9627 | 55.771 |
| Full Displacement | -2.6693 | 47.341 |

The amount of thrust that must allocated to each waterjet can be determined by solving (15) to find

$$T_{port} = \tau_x/2 + \tau_z/B \text{ and } T_{stbd} = \tau_x/2 - \tau_z/B. \tag{16}$$

The thrust developed on each waterjet is modeled using the linear relation between commanded waterjet motor speed and thrust output from equation (1).

*B. Simulation Results*

In order to determine the accuracy of the mathematical model presented above and to determine the hydrodynamic coefficients in the added mass matrix (7), Coriolis matrix (8), and drag matrices (10), (11), the experimental zig-zag and acceleration data are overlaid with simulation results for matching motor input. The values of the hydrodynamic coefficients were manually adjusted until the simulation results qualitatively matched the experimental data.

For the lightship condition, the straight-line acceleration data were compared against simulation results for three separate waterjet operating speeds: $RPM = \{0.8RPM_{max}, 0.9RPM_{max}, RPM_{max}\}$. Fig. 7-Fig. 10 display final results of the surge drag model in the vehicle's equations of motion. Specifically, the parameters that were identified to fit the experimental data were: $X_{\dot{u}}$, $X_u$, and $X_{u|u|}$. $X_u$ and $X_{u|u|}$ characterize the steady state surge velocity of the vehicle during the acceleration test (the flat area of the curves between 20 and 60 seconds on the figures below). Each steady state surge speed for the three test cases was tabulated and the results were fitted to a quadratic polynomial, as explained in Section IV.A. The equations displayed in Fig. 3 are of the form:

$$X_{drag} = D_{port} + D_{stbd} = X_{u|u|}u|u| + X_u u, \tag{17}$$

where $X_{drag}$ denotes drag in the vessel's surge direction. As the submerged hull form and wetted surface area varies with displacement, $X_u$ and $X_{u|u|}$ also vary with displacement. The term $X_u$ affects the transient behavior during acceleration/deceleration, i.e. between 0-10 seconds/past 60 seconds. $X_{\dot{u}}$ affects the slope of the simulation surge velocity curve during those beginning and end time periods.

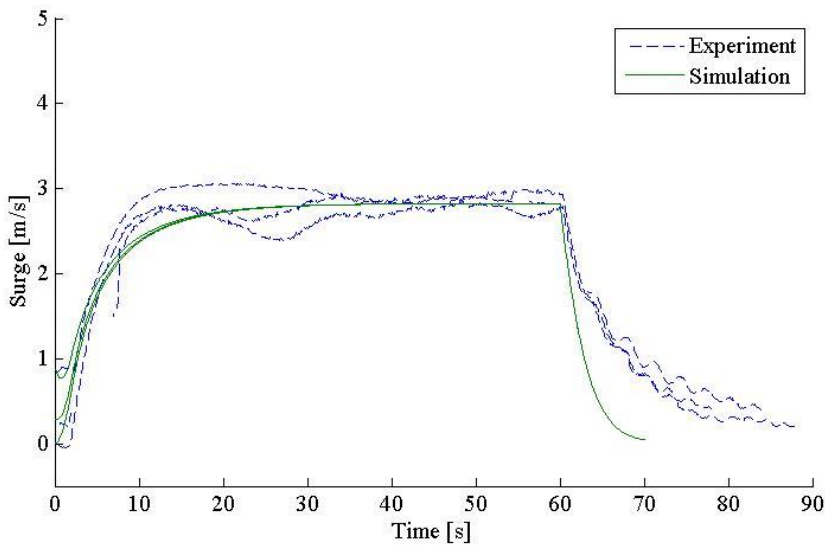

Fig. 7. Lightship surge histories for full throttle acceleration tests. Three experimental tests are shown, each at different initial conditions: $u$ = 0, 0.3, 0.9 m/s. Three simulation results are overlaid with initial conditions corresponding to field tests.

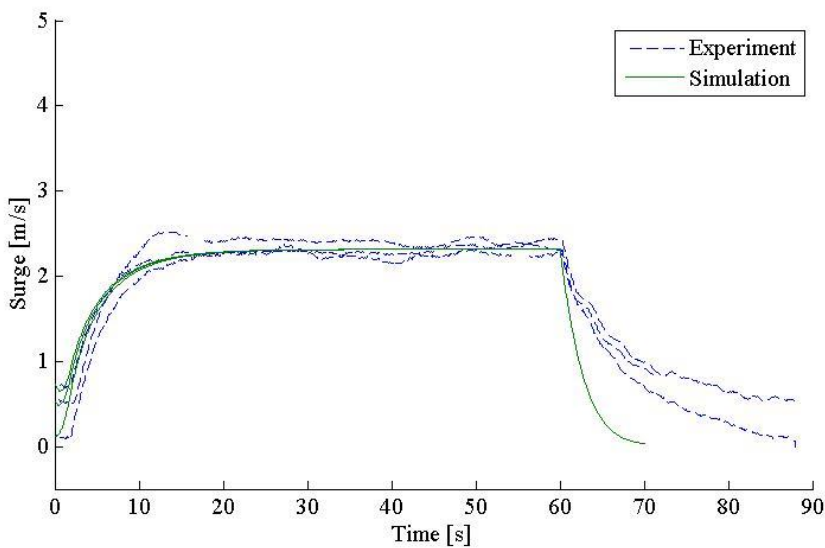

Fig. 8. Lightship surge histories for 90% throttle acceleration tests. Three experimental tests are shown, each at different initial conditions: $u$ = 0.13, 0.54, 0.75 m/s. Three simulation results are overlaid with initial conditions corresponding to field tests.

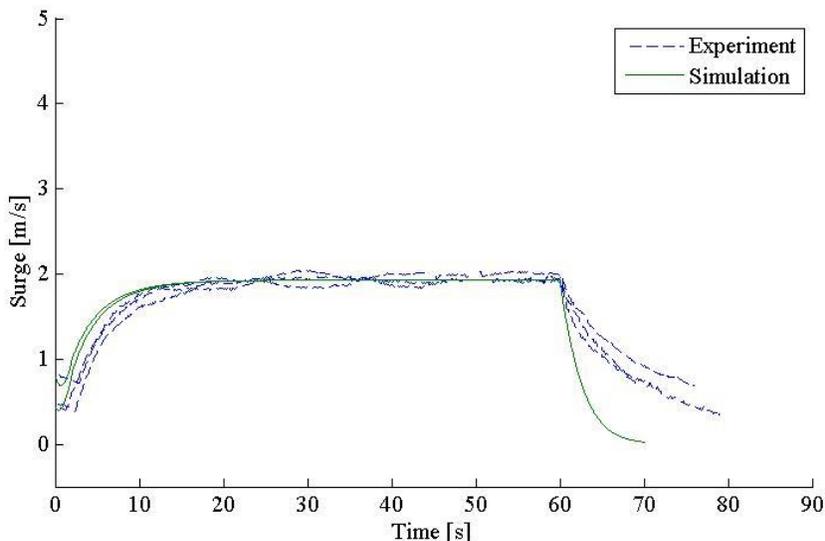

Fig. 9. Lightship surge histories for 80% throttle acceleration tests. Three experimental tests are shown, each at different initial conditions: $u = 0.44, 0.8, 0.8$ m/s. Two simulation results are overlaid with initial conditions corresponding to the field tests.

Results for the full displacement acceleration tests are displayed in Fig. 10-Fig. 14. When comparing the results, which cover the upper 40% of the vessel's motor operating limits and the two displacement conditions of interest in the ALR mission, one can conclude that the surge model of the WAM-V USV14 is adequately identified in the case of straight line forward motion.

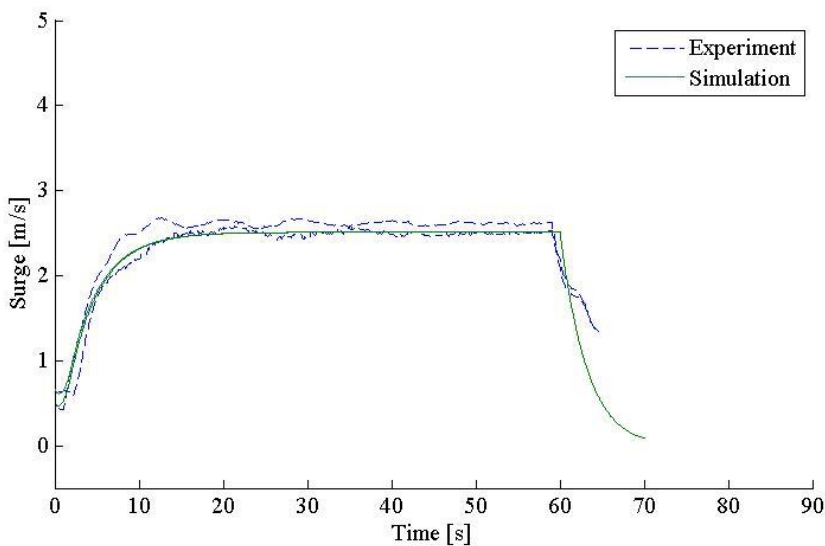

Fig. 10. Full displacement surge histories for full throttle acceleration tests. Two experimental tests are shown, each at different initial conditions: $u=0.5, 0.67$ m/s. Two simulation results are overlaid with corresponding initial conditions.

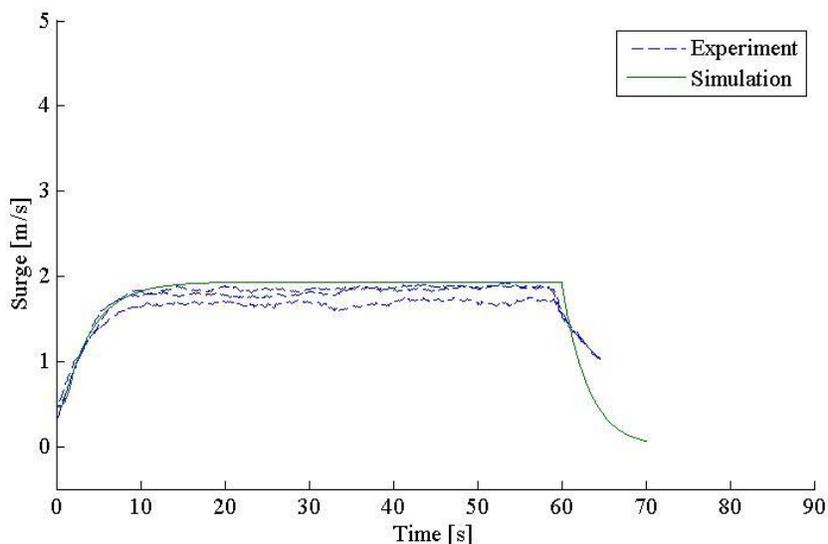

Fig. 11. Full displacement surge histories for 80% throttle acceleration tests. Two experimental tests are shown, each at the initial condition: $u$=0.5 m/s. One simulation result is overlaid with the same initial condition.

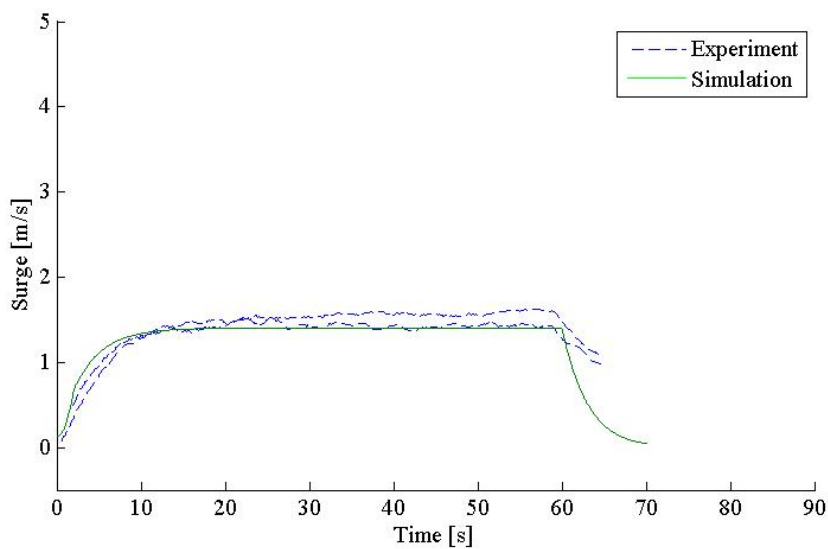

Fig. 12. Full displacement surge histories for 60% throttle acceleration tests. Two experimental tests are shown, each at the initial condition: $u$=0.13 m/s. One simulation result is overlaid with the same initial condition.

It can be noted that in Figs. 7-12 the experimental data show the boat accelerating faster when thrust is applied than it decelerates when thrust is removed. However, the simulations show the boat decelerating about as quickly as it accelerates. The reason for this is because, when moving, the real vehicle creates a wake of fluid traveling in the same direction as the vehicle. When thrust is removed and the vehicle starts to decelerate, the inertia of the fluid in the wake causes the fluid to pile up against the stern, pushing against the vehicle in the forward direction and slowing its deceleration.

Unfortunately, a simple quadratic drag model cannot capture this wake effect. Instead the relative rates of acceleration and deceleration are governed by the relative magnitude of the thrust and surge speed prior to acceleration/deceleration. One can see this from the equation of motion in the surge direction, which can be simplified as:

$$m\dot{u} = T - \frac{1}{2}\rho u^2 C_D A_f,$$

where $T$ is the constant thrust applied, $C_D$ is a drag coefficient and $A_f$ is the frontal area of the hulls. This can be rewritten as

$$\dot{u} = \frac{T}{m} - \frac{b}{m}u^2.$$

When the vehicle reaches a steady state speed $u_{ss}$, the acceleration is zero it can be seen that

$$u_{ss} = \sqrt{\frac{T}{b}}.$$

When thrust is initially applied, the acceleration is

$$\dot{u} = \frac{T}{m} - \frac{b}{m}u_0^2,$$

where $u_0$ is the initial velocity. As the surge speed increases the acceleration decreases to zero. After the vehicle has reached a steady state speed, at the instant that thrust is removed, the deceleration is

$$\dot{u} = -\frac{b}{m}u_{ss}^2 = -\frac{T}{m}.$$

Thus, one can see that, for the model, the magnitude of the initial acceleration is smaller than that of the initial deceleration when the model has some initial velocity. It may be possible circumvent this issue by including a switching added mass term in the surge equation of motion to produce the desired result. Since the bow is finer than the transom, one would expect the added mass associated with acceleration $m_{a1}$ to be smaller than that of deceleration $m_{a2}$. One could implement an equation, such as:

$$m_a = \frac{1 + \text{sgn}(\dot{u})}{2}m_{a1} + \frac{1 - \text{sgn}(\dot{u})}{2}m_{a2},$$

so that

$$m_a = \begin{cases} m_{a1}, & \dot{u} > 0 \\ m_{a2}, & \dot{u} < 0 \end{cases}$$

A plot of acceleration and deceleration shows that this could produce a result similar to what would be expected (Fig. 13). However, it is conventional to use a single added mass coefficient for both acceleration and deceleration. From a practical standpoint, it does not appear to be critical for the purposes of controller design, which is the main focus of the current study.

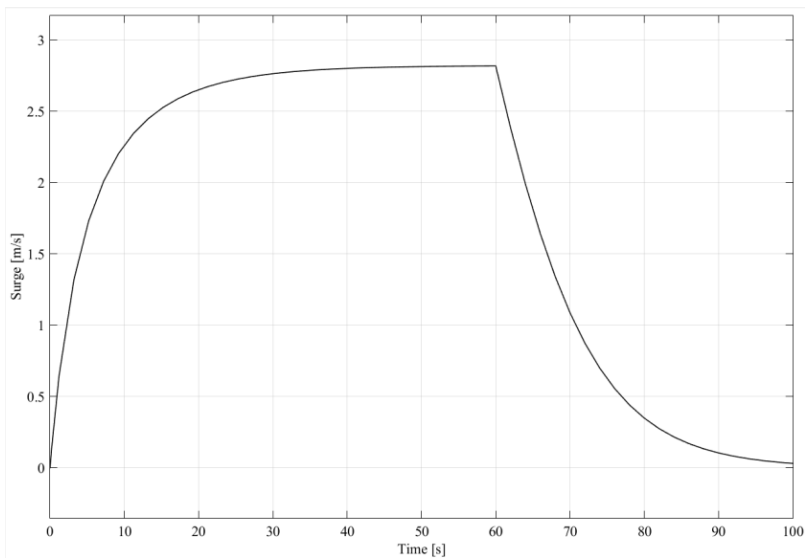

Fig. 13. Acceleration and deceleration with switching added mass term. The added mass for deceleration was made significantly larger than that of acceleration.

Now that verification of the model's straight line dynamics have been presented, the turning dynamics are addressed by examining the results of simulated zig-zag maneuvers. Fig. 14 shows an example comparison between the measured and simulated yaw rate for the zig-zag maneuver at 100% throttle. Based on the results of this and other yaw rate experiment/simulation comparisons, it was concluded that the dynamic model captures the turning dynamics of the vessel sufficiently well to be used for the design of a speed and heading controller for the ALR mission.

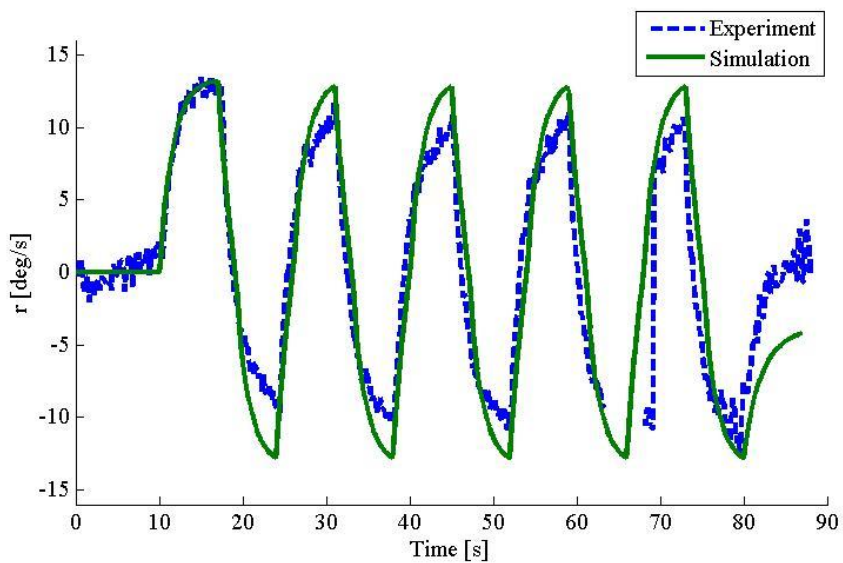

Fig. 14. Yaw rate data for the lightship 100% throttle zig-zag maneuver. Maximum experimental and simulated yaw rate is observed to be 13.2 deg/s.

Several assumptions are carried forward into the control system design process. First, it is assumed that the speed of the vessel is accurately predicted during straight line transits (i.e. when heading error has stabilized around zero) and that the surge velocity will be underestimated when heading error is large and hence the vehicle will turn aggressively to minimize the heading error. The second assumption about the model is that heading is accurately determined. Considering that the intent of this project is the development and implementation of an appropriate controller for the ALR mission, knowing that the speed and heading of the dynamic model can be trusted, or its errors anticipated, allows the search for a more accurate model to stop. The final values for the hydrodynamic coefficients used in the model are given in TABLE III. The middle column of this table contains the non-dimensional factors that are applied to the dimensional terms derived using strip theory and principles of drag for cylinders $(N_r, Y_r, Y_v)$. They are the correction factors for the disparity between theoretical vehicle dynamics and experimental data. In TABLE III: $\rho$ is water density, $T$ is the draft, $C_d$ is the drag coefficient for a lateral cylinder ($C_d = 1.1$), and $B_{hull}$ is the beam of the individual pontoon hulls.

TABLE III

HYDRODYNAMIC COEFFICIENTS FOR THE WAM-V USV14

| Coefficient Name | Non-Dimensional Factor | Dimensional Term |
|:---:|:---:|:---:|
| $N_{\dot{v}}$ | 2.5 | $-\pi\rho T^2 \dfrac{[(L-LCG)^2 + LCG^2]}{2}$ |
| $N_{\dot{r}}$ | 1.2 | $-\left\{\dfrac{4.75}{2}\pi\rho\dfrac{B_{hull}}{2}T^4 + \pi\rho T^2 \dfrac{[(L-LCG)^3 + LCG^3]}{3}\right\}$ |
| $X_{\dot{u}}$ | 0.075 | $-m$ |
| $Y_{\dot{r}}$ | 0.2 | $-\pi\rho T^2 \dfrac{[(L-LCG)^2 + LCG^2]}{2}$ |
| $Y_{\dot{v}}$ | 0.9 | $-\pi\rho T^2 L$ |
| $X_u$ | | See Fig. 3. |
| $Y_v$ | 0.5 | $-40\rho|v|\left[1.1 + 0.0045\dfrac{L}{T} - 0.1\dfrac{B_{hull}}{T} + 0.016\left(\dfrac{B_{hull}}{T}\right)^2\right]\left(\dfrac{\pi T L}{2}\right)$ |
| $N_r$ | 0.02 | $-\pi\rho\sqrt{(u^2 + v^2)}T^2 L^2$ |
| $N_v$ | 0.06 | $-\pi\rho\sqrt{(u^2 + v^2)}T^2 L$ |
| $Y_r$ | 6 | $-\pi\rho\sqrt{(u^2 + v^2)}T^2 L$ |
| $X_{u|u|}$ | | See Fig. 3. |
| $Y_{v|v|}$ | 1 | $-\rho T C_d L$ |
| $Y_{v|r|}$ | 1 | $-\rho T \dfrac{1.1}{2}[(L-LCG)^2 - LCG^2]$ |
| $Y_{r|v|}$ | 1 | Same as $Y_{v|r|}$ |
| $Y_{r|r|}$ | 1 | $-\rho T \dfrac{C_d}{3}[(L-LCG)^3 + LCG^3]$ |

| | | |
|---|---|---|
| $N_{v\|v\|}$ | 1 | Same as $Y_{r\|r\|}$ |
| $N_{v\|r\|}$ | 1 | Same as $Y_{r\|r\|}$ |
| $N_{r\|v\|}$ | 1 | Same as $Y_{r\|r\|}$ |
| $N_{r\|r\|}$ | 1 | $-\rho T \dfrac{C_d}{4}[(L-LCG)^4 + LCG^4]$ |

## VI. SPEED AND HEADING CONTROL

Due to the underactuation of the USV14, the lowest level of motion control is in surge speed and heading angle. This approach has worked well with this type of vehicle [3], [7],[8]. Several controllers regulating those motion primitives have been developed and implemented on the USV14.

### A. Backstepping Control

A surge speed and heading controller has been developed using feedback linearization [37]. The controller is the same as that presented by Liao et al. [24], apart from changes in formulating the controller for tracking control, rather than set point control, and in the use of a nonlinear drag model. Note that the approach is very similar to that developed by Ghommam et al. [17] for position and heading control of an underactuated surface vessel. As in both [17] and [24], the procedures developed by Panteley and Loria [28] can be used to show that the cascaded system is globally uniformly asymptotically stable.

To develop a surge speed controller, the surge equations of motion given by (3), (7)-(11), and (17) were reduced to:

$$(m - X_{\dot{u}})\dot{u} - (m - Y_{\dot{v}})vr - X_{drag} = \tau_x, \tag{18}$$

where $\tau_x$ is the surge component of the forces/moments (15) generated by the waterjet. Input state linearization [37] can be used to find a nonlinear feedback control law for the surge subsystem. Let $\xi_u$ be a linearizing control, related to the commanded value of $\tau_x$ from the controller by

$$\tau_x = (m - X_{\dot{u}})\xi_u - (m - Y_{\dot{v}})vr - X_{drag}. \tag{19}$$

With this value of $\tau_x$, (19) reduces to $\dot{u} = \xi_u$. If the surge tracking error is defined as $e_u = (u - u_d)$, where $u_d$ is the desired surge speed, and a linearizing control of the form $\xi_u = \dot{u}_d - k_u e_u$ is selected, the surge tracking error dynamics will be:

$$\dot{e}_u + k_u e_u = 0. \tag{20}$$

For positive values of the surge velocity gain $k_u$, the error dynamics for $e_u$ are stable. As in [24], a hyperbolic tangent form can be selected for the desired surge acceleration as

$$\dot{u}_d = \dot{u}_{a,max} \tanh[k_{a,max}(u_{d,ref} - u)/\dot{u}_{a,max}], \tag{21}$$

where $\dot{u}_{a,max}$ is the maximum allowed surge acceleration, $k_{a,max}$ is a positive surge acceleration gain that shapes the approach to the desired speed, and $u_{d,ref}$ is defined as

$$u_{d,ref} \equiv \min\{[u_{d,yaw} + (u_d - u_{d,yaw})\exp(-5.73|e_\psi|)], u_d\}, \tag{22}$$

where $e_\psi = \psi - \psi_d$ is the vehicle heading error, and $u_{d,yaw}$ is the desired surge velocity for periods of yaw motion. The function min{ } denotes taking the minimum of the two values in the function's argument. The intent of including the $u_{d,ref}$ term is to keep the surge speed low during turning maneuvers in order to minimize the vehicle's turning radius.

To develop a heading controller, the yaw rate equations of motion given by (3), and (7)-(11) were simplified as:

$$(I_z - N_{\dot{r}})\dot{r} - (-X_{\dot{u}} + Y_{\dot{v}})uv - N_r r = \tau_z. \tag{23}$$

As above, input state linearization was used to identify a nonlinear feedback control law for the yaw subsystem. Let $\xi_\psi$ be a linearizing control related to the commanded torque from the controller $\tau_z$ by the relation

$$\tau_z = (I_z - N_{\dot{r}})\xi_\psi - (-X_{\dot{u}} + Y_{\dot{v}})uv - N_r r. \tag{24}$$

Using this in (23) gives $\dot{r} = \ddot{\psi} = \xi_\psi$. If the heading tracking error is defined as $e_\psi = (\psi - \psi_d)$, where $\psi_d$ is the desired heading angle, and an input of the form $\xi_\psi = \ddot{\psi} = \ddot{\psi}_d - k_1 e_\psi - k_2 \dot{e}_\psi$ is selected, the error dynamics will be:

$$\ddot{e}_\psi + k_2 \dot{e}_\psi + k_1 e_\psi = 0, \tag{25}$$

which are exponentially stable when the control gains are positive: $k_1 > 0$ and $k_2 > 0$. As only the desired yaw $\psi_d$, and not also $\dot{\psi}_d$ and $\ddot{\psi}_d$, is provided by the ALR system, the control law reduces to:

$$\tau_z = (I_z - N_{\dot{r}})(-k_1 e_\psi - k_2 r) - (-X_{\dot{u}} + Y_{\dot{v}})uv - N_r r. \tag{26}$$

As shown in [24], the combined surge speed and heading control laws make the system state $(u, r, \psi)^T$ bounded and globally uniformly asymptotically convergent to $(u_d, 0, \psi_d)^T$ when $t \to \infty$. They are also well-defined and bounded for all $t > 0$ when $\{k_u, k_1, k_2\} > 0$.

Examining the sway velocity $v$ subsystem,

$$(m - Y_{\dot{v}})\dot{v} + (m - X_{\dot{u}})ur - Y_v v = 0, \tag{27}$$

And using the error dynamics produced in (20) and (25), one can readily see that the sway subsystem in (27) reduces to,

$$(m - Y_{\dot{v}})\dot{v} = -(m - X_{\dot{u}})u_d \dot{\psi}_d + Y_v v, \tag{28}$$

when $t \to \infty$, which is exponentially stable for the setpoint case of $\dot{\psi}_d = 0$. A block diagram of the controller is shown in Fig. 15.

                                                    21

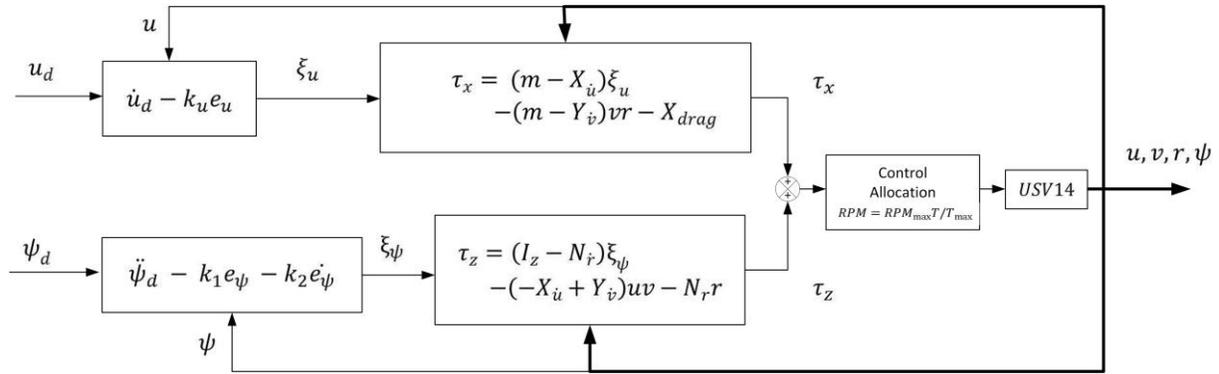

Fig. 15: Structure of the nonlinear backstepping controller.

During initial on-water controller testing it was found that the waterjet motors were exhibiting a chatter-like behavior. An examination of the thrust model Section IV.A shows that the commanded speed of each waterjet is sensitive to small changes in desired thrust. To reduce the chattering effect, the commanded surge force (19) is scaled by 0.5 and the commanded yaw moment (26) is scaled by 0.05. Coupled with the fact that the waterjet pump motor driver has a maximum speed that it can output, the scale factors basically act like a saturation function for the commanded motor speed to alleviate the chattering issue.

Despite the fact that a reduced set of the equations of motion in equations (3)-(11) were used to develop the backstepping surge speed and heading controllers, the backstepping controller was found to be robust to unmodeled dynamics throughout the range of the USV operating conditions tested.

### B. Backstepping Yaw Controller with Adaptive Speed Control

Based on the results of simulation and initial on-water testing of the non-adaptive backstepping controller, the need for adaptive speed control was established. Although the change in turning dynamics between displacement conditions does not appear to drastically affect heading control performance, the change in drag significantly affects the performance of surge speed control. To remedy the issue of steady state speed errors fluctuating with mass and drag parameters throughout the ALR mission, an adaptive surge speed control law was developed and implemented in combination with the backstepping heading controller developed in Section VI.A above.

By combining (17) and (18), one can write the surge equation of motion as:

$$h\dot{u} - (m - Y_{\dot{v}})vr - X_u u - X_{u|u|}u|u| = \tau_x, \tag{29}$$

where $h \equiv (m - X_{\dot{u}})$. A control law for the thrust commanded by the controller of the form

$$\tau_x = -(m - Y_{\dot{v}})vr - \hat{X}_u u - \hat{X}_{u|u|}u|u| + \hat{a}_d u_d, \tag{30}$$

where $\hat{X}_u$, $\hat{X}_{u|u|}$ and $\hat{a}_d$ are estimated parameters, is chosen to decouple the surge dynamics from those of sway and yaw. With this control law, (29) becomes

$$h\dot{u} = (X_u - \hat{X}_u)u + (X_{u|u|} - \hat{X}_{u|u|})u|u| + \hat{a}_d u_d. \tag{31}$$

A linear reference model of the form

$$\dot{u}_m = -a_m u_m + b_m u_d, \tag{32}$$

is chosen and the model error is defined as $e_m \equiv (u - u_m)$. With this reference model (31) can be reduced to an equation describing the surge speed error dynamics:

$$h\dot{e}_m + h a_m e_m = (h a_m + \tilde{a}_1)u + \tilde{a}_2 u|u| - (h b_m - \hat{a}_d)u_d, \tag{33}$$

where $\tilde{a}_1 \equiv (X_u - \hat{X}_u)$ and $\tilde{a}_2 \equiv (X_{u|u|} - \hat{X}_{u|u|})$ are parameter estimation errors. An adaptation law of the form

$$\left. \begin{array}{lll} \frac{d\tilde{a}_1}{dt} & = -\frac{d\hat{X}_u}{dt} = & -\gamma e_m u \, \mathrm{sgn}(h) \\ \frac{d\tilde{a}_2}{dt} & = -\frac{d\hat{X}_{u|u|}}{dt} = & -\gamma e_m u|u| \, \mathrm{sgn}(h) \\ \frac{d\hat{a}_d}{dt} & = & -\gamma e_m u_d \, \mathrm{sgn}(h) \end{array} \right\}, \tag{34}$$

is chosen for stability, where $\mathrm{sgn}(h)$ is the sign (signum) function and $\gamma$ is a positive adaptation gain. The stability of the surge speed controller can be shown for the Lyapunov function candidate:

$$V = |h|e_m^2 + \frac{1}{\gamma}[(h a_m + \tilde{a}_1)^2 + \tilde{a}_2^2 + (h b_m - \hat{a}_d)^2]. \tag{35}$$

Using $|h| = h \, \mathrm{sgn}(h)$ together with (33) and (34), the derivative of the Lyapunov function reduces to

$$\dot{V} = -2a_m e_m^2 |h|, \tag{36}$$

which is negative definite for $a_m > 0$. The coefficients $a_m$ and $b_m$ of the reference model (32) can be determined by linearizing the hyperbolic tangent based model acceleration for the nonlinear backstepping controller (21) with $u_{d,ref} = u_d$. When the surge speed $u$ achieved by the USV approaches that of the reference model $u_m$, the argument of the hyperbolic tangent function becomes small and the model acceleration can be approximated as:

$$\dot{u}_m \approx \frac{\dot{u}_{a,max} k_{a,max}}{u_{a,max}}(u_d - u_m), \tag{37}$$

such that $a_m = b_m = \dot{u}_{a,max} k_{a,max}/u_{a,max}$. A block diagram of the backstepping adaptive controller is shown in Fig. 16.

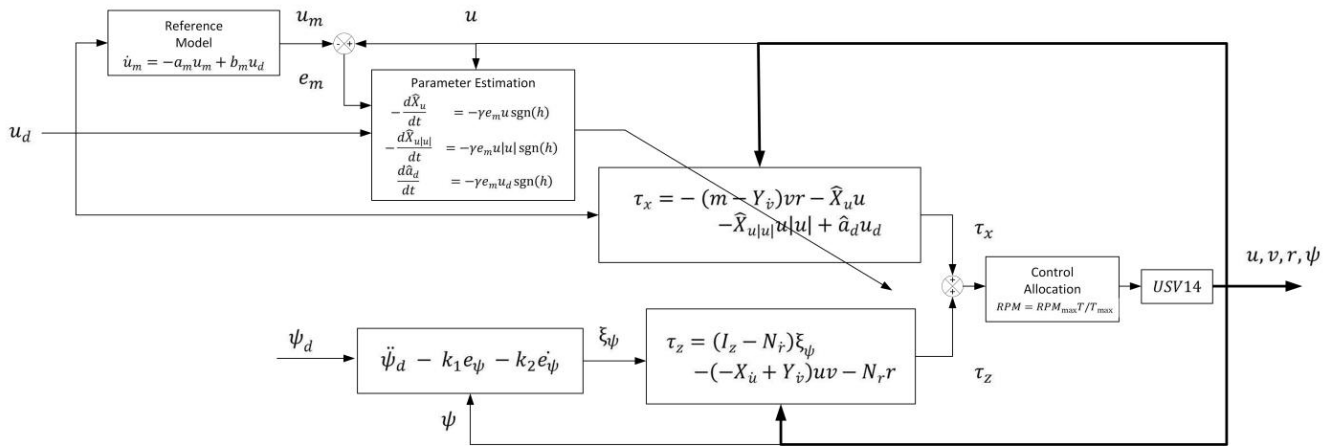

Fig. 16: Structure of the adaptive backstepping controller.

As in Section VI.A, the commanded surge force $\tau_x$ is multiplied by a factor of 0.5 to prevent chattering. During field trials the adaptive backstepping controller was found to work well with large, sudden changes in payload mass (see Section VII). In practice, it was also found that when the parameter estimates $\hat{X}_{u|u|}$ and $\hat{a}_d$ were held constant, the adaptive controller still performed well, because the linear drag term in the control law (30) dominates the other two terms at low speeds ($u \leq 1.5$ m/s).

*C. Comparative Set Point Tests*

In order to establish a reasonable comparison between the two controllers, a set point test is conducted on each controller. The test first consists of placing the USV in a predetermined position at rest with a predetermined initial heading. The test is started by delivering a constant speed command and an initial desired heading command that is the same as the initial heading of the USV. The USV accelerates towards the desired speed and also attempts to maintain heading. At a specified time (15 seconds), the heading command is changed to a value 90 degrees from the initial heading and the USV executes a turn towards the new desired heading. The metrics for evaluating controller performance during this test are steady state speed and heading error before and after the jump in heading command. The test is conducted a total of three times for each controller and the performance metrics are averaged over all three tests.

To quantify the difference in performance between two controllers, the difference between the errors is divided by the larger error, resulting in the percentage amount that the lower error is 'better' when compared to the larger. This metric is relative to the least desirable result, which is the baseline for comparison. For example, consider results $A$ and $B$ where $A$ is more desirable than $B$. The difference between $A$ and $B$ is divided by $B$ and the result $\lambda = (B - A)/\max(A, B) \times 100\%$ is said to show that "$A$ performs $(B - A)/\max(A, B) \times 100\%$ better than $B$." Since the metric we are comparing is steady state error, a positive value for $\lambda$ denotes an improvement in $A$ with respect to $B$. Naturally, a negative value would denote a performance decrease. This convention applies for all subsequent sections where two control results are compared.

For the results presented in this section, the inputs used for the test are $\psi_d = \psi(t = 0)$ for $0 \leq t \leq 15$ seconds, $\psi_d = \psi(t = 0) +$ 90 degrees for $t > 15$ seconds, and $u_d = 1.5$ m/s. For the non-adaptive backstepping controller the control gains are set to $k_1 = 0.1, k_2 = 1.0, k_u = 8.0$, and $k_{a,max} = 1.2$. The adaptive backstepping controller has control gains $k_1 = 0.1, k_2 = 1.0$, $k_{a,max} = 1.2$, and $\gamma = 0.05$. The backstepping controller gains were found (through previous testing) to work well on the USV. The WAM-V USV14 was in the lightship condition for the tests, a 10-15 knot sustained wind blew from the southeast (approximately $160^0$), and a slight tidal current of <0.2 knots flowed Northerly (approximately $345^0$). The small wind driven waves were taken to be negligible in affecting the performance of the vehicle.

The results of one set point test when the USV is controlled by the non-adaptive backstepping controller are shown in Fig. 17. The average speed error at steady state before the jump in heading is 0.25 m/s, and decreases to 0.24 m/s after the jump. The average heading error at steady state before the jump is 0.74 degrees and decreases to 0.41 degrees afterwards.

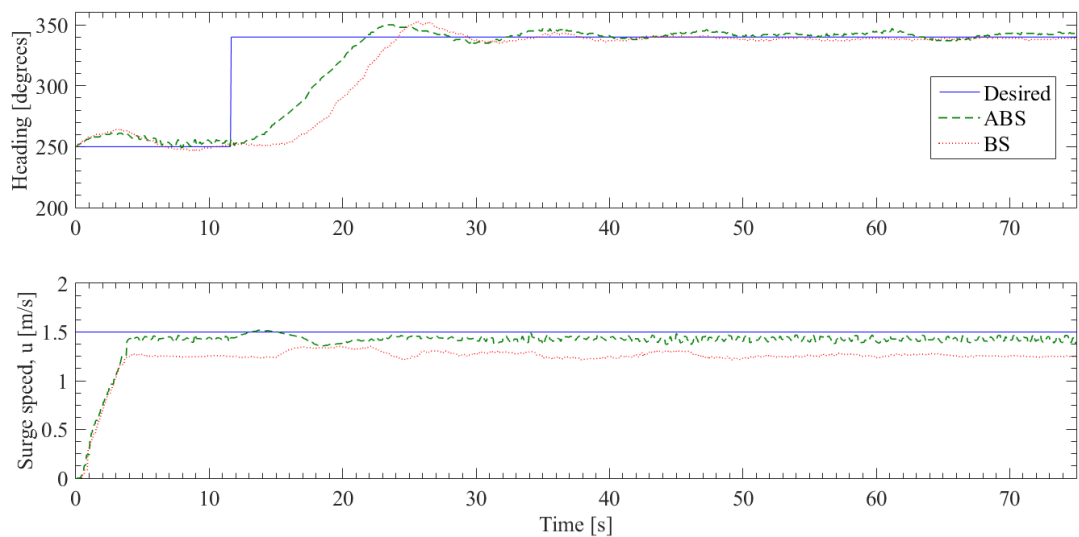

Fig. 17. Results for the set point test using the backstepping (BS) and adaptive backstepping (ABS) controllers.

Results for the same set point test using the adaptive backstepping controller are overlaid on those of the backstepping controller in Fig. 16 above. The average speed error at steady state before the jump in heading is 0.065 m/s, and 0.070 m/s after the jump. The average heading error at steady state before and after the jump is 1.9 degrees.

The overall results averaged over three runs for each controller are shown in Table IV. The controllers have been annotated as ABS and BS adaptive backstepping and backstepping, respectively. For brevity, the charts only show the comparisons of the controllers after the jump in heading command. The results show that the non-adaptive backstepping controller shows 62% better heading performance over the adaptive controller before the heading change. After the heading change, the non-adaptive backstepping controller shows a 78% better heading performance than the adaptive controller.

| Controller: | ABS | BS |
|---|---|---|
| Steady State Speed Error Before Jump (m/s): | 0.065 | 0.25 |
| Percent Error: | 4% | 17% |
| Steady State Speed Error After Jump (m/s): | 0.070 | 0.24 |
| Percent Error: | 5% | 16% |
| Steady State Heading Error Before Jump (deg): | 1.9 | 0.74 |
| Steady State Heading Error After Jump (deg): | 1.9 | 0.41 |

## VII. VARIABLE MASS AND DRAG EXPERIMENTS

Mock ALR tests were performed in Dania Beach and Hollywood, Florida USA. In order for the vehicle to operate effectively, it cannot be used in sea state greater than 1 or winds higher than 15 knots. The maximum sea state and wind conditions for operations were determined by trial and error.

During the experiments the vehicle was driven under closed loop speed and heading control in the ICW. Vehicle heading, acceleration, position, speed over ground, and motor command data were recorded, as were the wind and current speed/direction, and wave characteristics. The WAM-V USV14 was tested over a range of surge speeds that are expected for the ALR mission.

### A. Variable Mass

One of the possible scenarios for the launch phase is the AUV being dropped into the water from the USV's undercarriage by the ALR system. In order to simulate this scenario, a test case where the USV's mass properties vary during operation is conducted. Specifically, the mass change test involves the USV beginning in the full displacement condition and coming to a set desired speed and heading. The mass and drag properties of the lightship condition are incorporated into the adaptive and non-adaptive backstepping control laws (creating the uncertain mass and drag situation). Once the USV reaches steady state in both speed and heading, an AUV-like mass is dropped from the payload and the USV continues along its course. No real-time feedback is provided to the low-level controller about the actual displacement condition during the test, either before or after the drop. The performance of each controller across the two displacement conditions is then compared against the other controllers.

For the experimental setup, the AUV-like mass is a Polyvinyl Chloride (PVC) pipe measuring 130 cm in length, 12 cm in diameter, weighted with lead shot and capped at both ends (Fig. 18a). The overall weight in air of the PVC apparatus is 39 kg (approximately the same as a REMUS 100 AUV), with the weight evenly distributed along the length. Four buoys (each rated at 13.6 kg buoyancy) are lashed to the PVC apparatus, two at each end. The device is secured to the undercarriage of the USV using both fore and aft lines. The aft line runs through a loop directly above the aft section of the PVC device and forward to a quick-release mechanism, where both lines are clipped in, thus suspending the PVC device under the payload tray of the USV and along the USV centerline. The quick release mechanism holds the two lines together until a pin is pulled, which releases the connection and the device falls into the water. A line is tied to the pin, which allows a user to pull the pin from a distance (from a chase boat in front of the USV during the test).

A view of the USV in the full displacement condition from below as it is lifted with a crane is shown in Fig. 18b. Notice that the AUV-like object is located along the centerline of the USV, and the two additional PVC pipes provide the weight of the intended ALR mechanical system (68 kg).

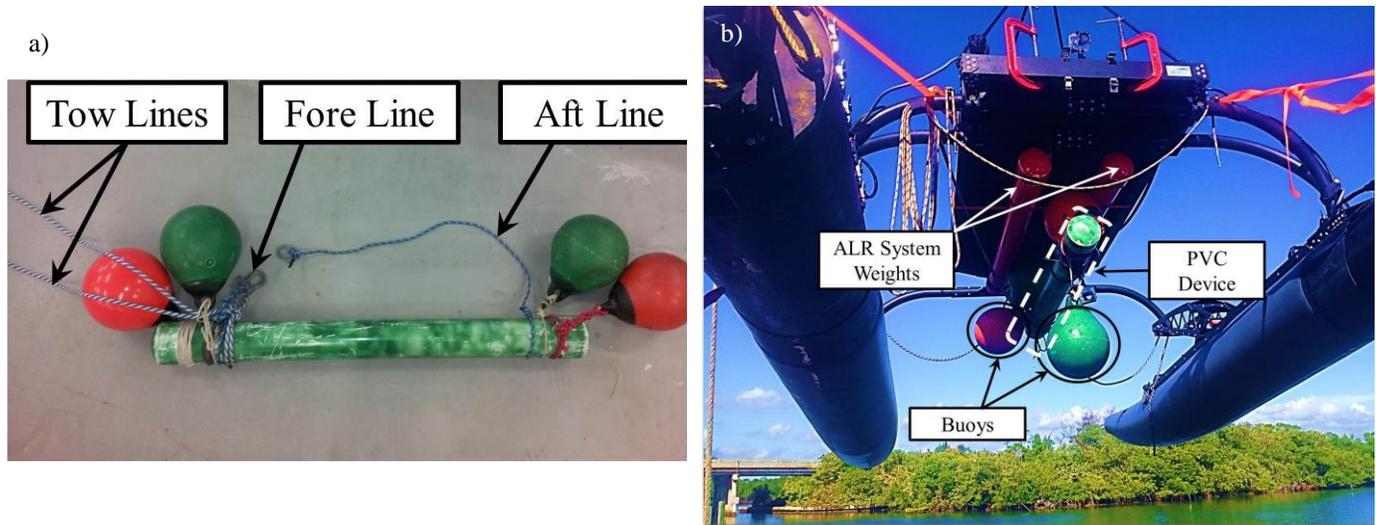

Fig. 18. a) PVC device used to mimic a REMUS 100 AUV. b) Bottom view of the WAM-V USV14 in the full displacement condition.

The results of the mass change test when the USV is controlled by the backstepping controller tuned to the lightship condition are shown in Fig. 18. For the time up to the drop of the PVC device, the USV is in the full displacement condition traveling along a desired heading of 150º at a desired speed of 1 m/s. The quick release mechanism can only be pulled from the front of the USV. Accordingly, the drop is marked by a spike in surge velocity (and sometimes a perturbation to the heading) which is partly due to the impulsive force applied to the quick release line. Because the impulsive force is unmeasured, the data corresponding to the time when the force is applied are not included in the steady state averages. A picture, taken just before the mass drop occurs, is shown in Fig. 19. The average speed error for the backstepping controller at steady state before the drop is 0.14 m/s, and decreases to 0.13 m/s after the drop. The data are shifted in time so that the results are synchronized with those of the adaptive backstepping controller presented below where the drop occurs at t = 78 seconds. The average heading error at steady state before the drop is 1.4 degrees and increases to 2.6 degrees afterwards.

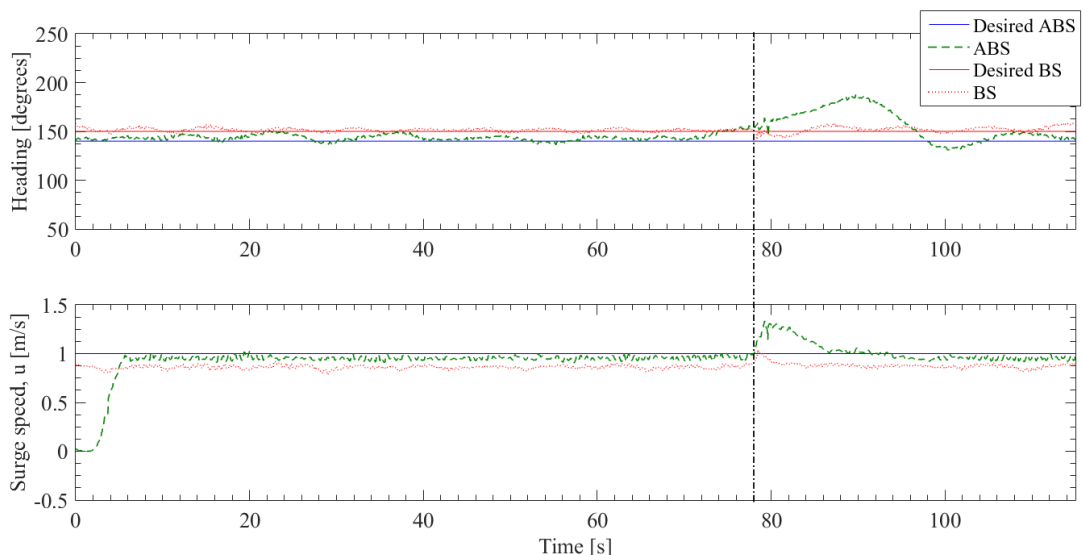

Fig. 19. Example results of the variable mass test using the backstepping controller and the adaptive backstepping controller. Data from the different tests are synched according to the time of the mass drop for the ABS controller. The large perturbations in surge speed and heading in the adaptive backstepping controller test at the time of the mass drop are attributed to a large impulse delivered through the quick release line.

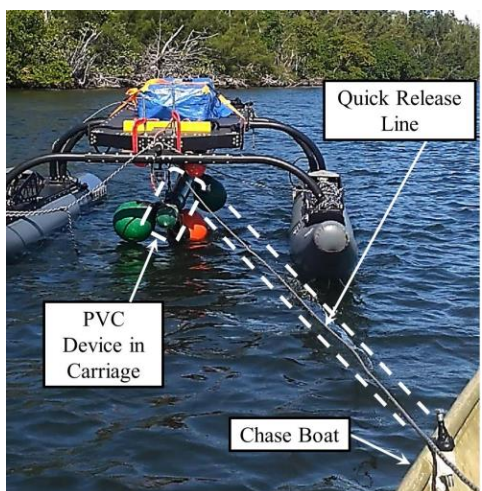

Fig. 20. Snapshot just before the PVC device is released from the USV undercarriage during the variable mass test.

Results for the variable mass test using the adaptive backstepping controller tuned to the lightship condition are also shown in Fig. 18 above. The average speed error at steady state before the drop is 0.046 m/s, and decreases to 0.042 m/s after the drop, which is synched to the backstepping controller such that it occurs at t = 78 seconds. The average heading error at steady state before the drop is 3.3 degrees and increases to 4.7 degrees afterwards.

A comparison of the speed and heading errors for the three controllers is shown in Table V. For brevity, the full displacement condition results are excluded from the comparison table. The results show that for each backstepping controller the speed error decreased when the USV transitioned from the full displacement to the lightship condition. The magnitude of the speed errors, however, show that the adaptive controller performs 67% better than the backstepping controller when in the full displacement condition and performs 68% better than the backstepping controller when in the lightship condition. Note that the short period of time during which the USV is reacting to the impulse of the quick-release mechanism are excluded from the analysis of the steady state characteristics of the controlled USV.

TABLE V

COMPARISON OF THE ADAPTIVE BACKSTEPPING AND BACKSTEPPING CONTROLLERS FOR THE VARIABLE MASS TEST. USING THE METRIC IDENTIFIED IN SECTION VI.C, THE STEADY STATE SPEED ERROR OF THE ABS CONTROLLER IS 68% BETTER THAN THE BS CONTROLLER AFTER THE DROP. HOWEVER, THE STEADY STATE HEADING ERROR OF THE BS CONTROLLER IS 43% BETTER THAN THE ABS CONTROLLER AFTER THE DROP.

| Controller: | ABS | BS |
|---|---|---|
| Steady State Speed Error Before Drop (m/s): | 0.05 | 0.14 |
| Percent Error: | 4.6% | 14% |
| Steady State Speed Error After Drop (m/s): | 0.04 | 0.13 |
| Percent Error: | 4% | 13% |
| Steady State Heading Error Before Drop (deg): | 3.3 | 1.3 |
| Steady State Heading Error After Drop (deg): | 3.8 | 1.4 |

The results for heading do not show the same trend in performance. The non-adaptive backstepping controller shows 61% better heading performance over the adaptive controller in the full displacement condition. For the lightship condition, the non-adaptive backstepping controller shows 43% better heading performance over the adaptive controller.

*B. Variable Drag*

A second scenario expected for the ALR mission is when the USV is navigating a straight trajectory in the lightship condition and the AUV (approaching from behind) impulsively docks with a recovery device suspended below the surface. This towing situation would create an abrupt change in drag experienced by the USV, and also alter the turning dynamics, i.e. a change in both the heading and surge subsystems of the USV would occur. As in the change of mass testing, it is assumed that no real-time feedback about the presence of the AUV is provided to the USV low-level controller, which is tuned for the lightship condition.

To create this scenario, the AUV-like PVC device was secured at two tow points on the aft section of the USV payload tray using a tow line. Each line is 7.3 m long, and attached to the front section of the PVC device. The PVC device is initially carried aboard a chase boat behind the USV so that the towing lines are slack. Once the USV comes to a steady heading and speed, the PVC device is dropped into the water from the chase boat and begins to be towed by the USV (Fig. 21). For this set of tests, since the mass of the USV is not intended to be altered, the PVC device was weighted to 28 kg, which eased the lifting of the device to and from the chase boat. The shape of the device was unaltered from the other tests; however the buoys sat higher in the water which decreased the overall drag of the device. In order to determine the amount of drag experienced by the USV during the variable drag tests, the PVC device is towed behind the chase boat. The tow line is attached to a scale, which is then attached to the tow point on the chase boat. The control box is used to record the boat's velocity and values are read from the scale for different speeds ranging from 0.5 to 1.6 m/s. Fig. 22 shows the drag model developed for the PVC device. Each point on the plot represents the average value of force recorded for that speed and the solid line is the curve fit to the points, valid for the range from 0.5 to 1.6 m/s. The data are averaged over a total of 76 force readings from the scale. For the data presented in this section, with a desired speed of 1.0 m/s, the drag of the PVC device is estimated at 84 N. The same drag test conducted for the REMUS 100 AUV resulted in a drag value of no more than 23 N at 1.0 m/s. The variable drag tests, therefore, allow for a towed recovery device drag of at least 61 N.

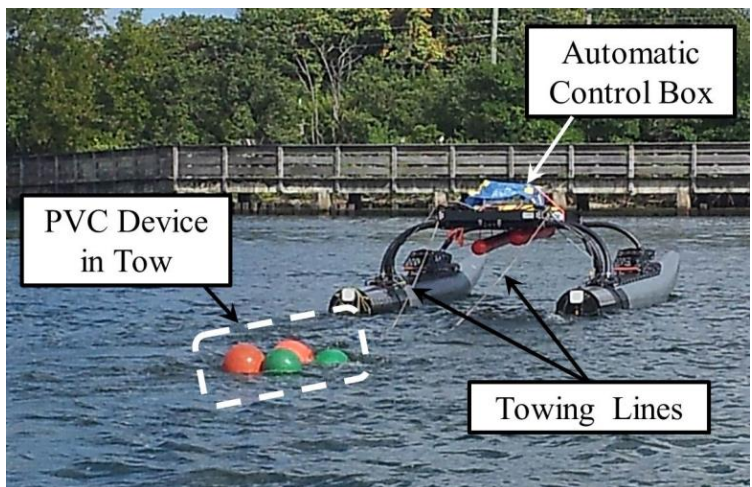

Fig. 21. Picture of the WAM-V USV14 towing the AUV-like object through the water during the variable drag testing.

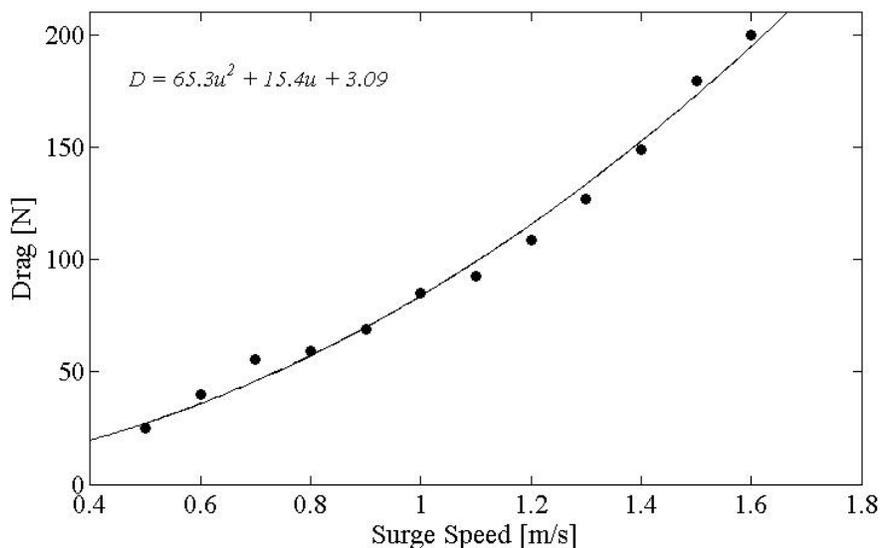

Fig. 22. Drag model for the 28 kg PVC device used in the variable drag tests.

Results of the variable drag test using the non-adaptive backstepping controller tuned to the lightship condition are shown in Fig.

23. The average speed error at steady state before the drop is 0.13 m/s, and increases to 0.17 m/s after the drop. The results are

shifted in time so that the drop occurs at t = 42 seconds and are synchronized with those of the adaptive backstepping controller.

The average heading error at steady state before the drop is 0.65 degrees and increases to 0.94 degrees afterwards. The same test was conducted for a desired speed of 1.2 m/s, however the drag of the AUV-like PVC device was too high and the propulsion system could not achieve the desired speed; the heading control also degraded significantly. Only the tests where desired speed was 1.0 m/s are shown. The quantitative results are averages of the data shown in the figures only, and are similar to the results from other tests.

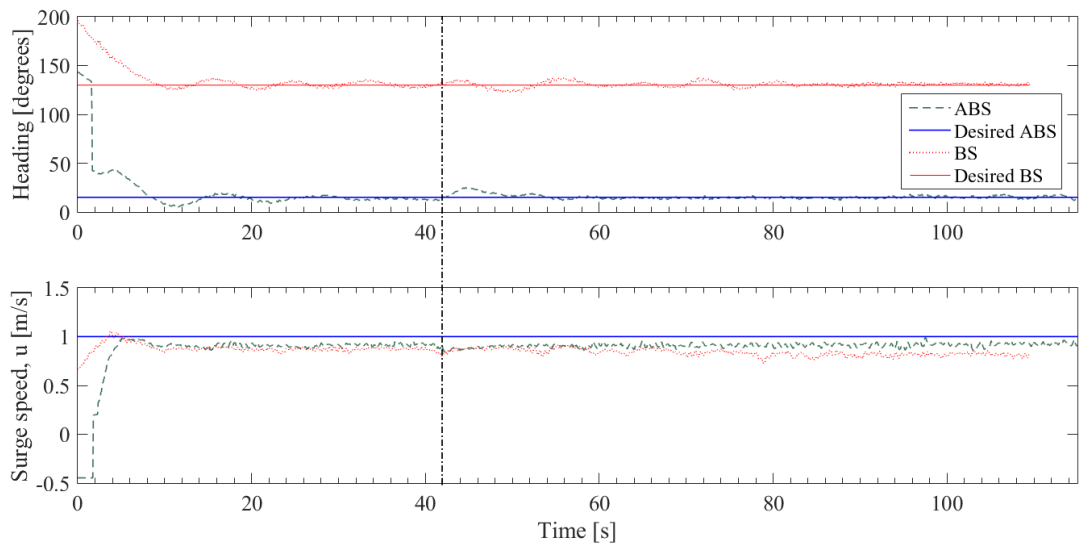

Fig. 23. Results for the variable drag test using the backstepping and the adaptive backstepping controller. The two tests are synched according to the time of the drop.

Results using the adaptive backstepping controller tuned initially to the lightship condition are also shown in Fig. 22. The average speed error at steady state before the drop is 0.091 m/s and after the drop it is 0.090 m/s. The drop occurs at t = 42 seconds. The average heading error at steady state before the drop is 1.81 degrees and decreases to 0.35 degrees afterwards.

A comparison of the speed and heading errors for the three controllers is shown in Table VI. For brevity, the results for the comparison chart before the drop are excluded. The results show that for each controller the speed error increased when the USV transitioned from the lightship to towing condition; however, the adaptive controller showed a 58% improvement on the backstepping controller while in the lightship displacement condition. The adaptive controller performs 64% better than the backstepping controller while in the towing condition.

TABLE VI
COMPARISON OF THE ADAPTIVE BACKSTEPPING AND BACKSTEPPING CONTROLLERS FOR THE VARIABLE DRAG TEST. USING THE METRIC IDENTIFIED IN SECTION VI.C, THE STEADY STATE SPEED ERROR OF THE ABS CONTROLLER IS 64% BETTER THAN THE BS CONTROLLER AFTER THE DROP. HOWEVER, THE STEADY STATE HEADING ERROR OF THE BS CONTROLLER IS 66% BETTER THAN THE ABS CONTROLLER AFTER THE DROP.

| Controller: | ABS | BS |
| --- | --- | --- |
| Steady State Speed Error Before Drop (m/s): | 0.06 | 0.13 |
| Percent Error: | 5.5% | 13% |
| Steady State Speed Error After Drop (m/s): | 0.06 | 0.18 |
| Percent Error: | 6% | 18% |
| Steady State Heading Error Before Drop (deg): | 0.1 | 0.5 |
| Steady State Heading Error After Drop (deg): | 2.6 | 0.9 |

The results for heading are as follows. The adaptive backstepping controller shows 83% better heading performance over the non-adaptive backstepping controller in the lightship condition. For the towing condition, the non-adaptive backstepping controller shows 66% better heading performance over the adaptive controller.

The maximum speed at which the WAM-V can maintain heading control while towing a device with 84 N of drag is 1.0 m/s. This number will decrease if the drag of the recovery device decreases because the initial estimate of the REMUS 100 drag at 1.0 m/s is 23 N – far less than the AUV-like object used for testing (see Fig. 22).

*C. Variable Mass and Drag*

A third scenario expected for the ALR mission is where the USV is navigating a straight trajectory in the full displacement condition when the AUV is lowered/dropped into the water and is still attached to the ALR device for some time before being completely launched from the USV. This carriage to towing transition would create a change in the drag and mass of the USV. The turning dynamics would also be altered as in the case of the variable drag testing presented in Section VII.B. As in the other testing, it is assumed that no real-time feedback about the presence of the AUV is provided to the USV low-level controller.

To conduct the variable mass and drag experiment, the setups for the variable mass (Section VII.A) and variable drag test (Section VII.B) are combined. The PVC device is loaded in the undercarriage of the USV and the tow lines are coiled on the payload tray and attached at the ends to the rear tow points on the USV. The USV comes to a steady heading and speed in the full displacement condition, the quick release is pulled, and the device falls into the water. As the USV continues forward the tow lines are brought into tension and the AUV is towed behind the USV.

Results of the variable mass and drag test using the non-adaptive backstepping controller tuned to the lightship condition are shown in Fig. 24. In this case, the PVC device separated cleanly from the suspension points when the quick-release was pulled so the device immediately transitioned to towing at distance. The average speed error at steady state before the drop is 0.14 m/s, and increases to 0.19 m/s after the drop. As above, the results of the backstepping controller test are synchronized with those of the adaptive backstepping controller, such that the drop occurs at t = 87 seconds. The average heading error at steady state before the drop is 0.36 degrees and increases to 2.9 degrees afterwards. Results using the adaptive backstepping controller tuned initially to the lightship condition are also shown in Fig. 23. The average speed error at steady state before the drop is 0.071 m/s and after the drop it is 0.066 m/s. The average heading error at steady state before the drop is 3.1 degrees and increases to 3.4 degrees afterwards.

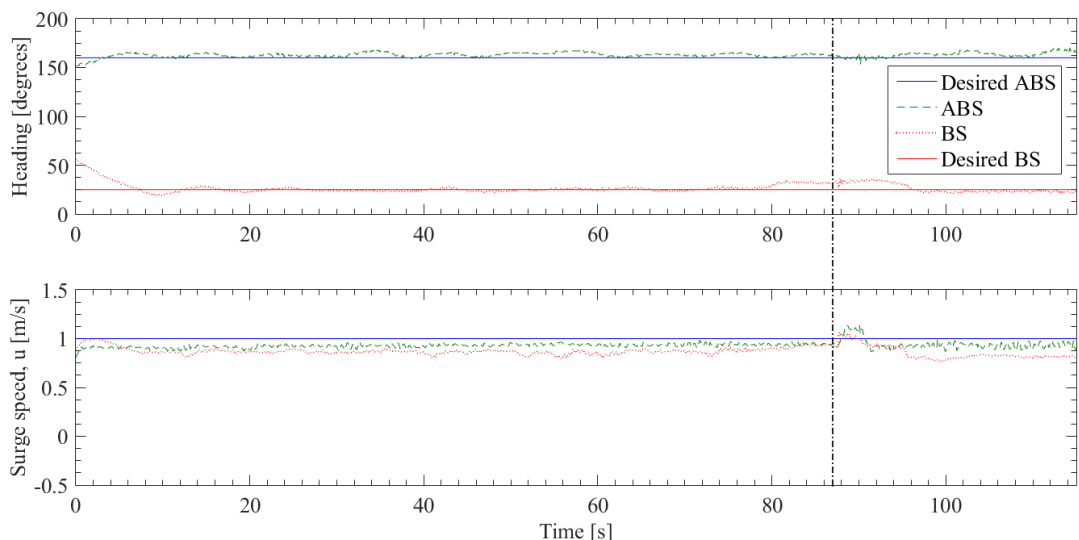

Fig. 24. Results of the variable mass and drag test using the backstepping and the adaptive backstepping controllers. The data from the different tests are synched according to the time of the mass drop.

A comparison of the speed and heading errors for the controllers is shown in Table VII. For brevity, the comparison chart excluded the results before the drop. The results show that for the non-adaptive controllers the speed error increased when the USV transitioned from the full displacement to towing condition. The magnitude of the speed errors show that the adaptive controller retains an advantage of 48% on the backstepping controller while in the full displacement condition. The adaptive controller performs 65% better than the backstepping controller while in the towing condition. Overall, the adaptive controller out-performed both non-adaptive controllers in in transitioning from full displacement to lightship condition.

The results for heading are as follows. The non-adaptive backstepping controller shows 89% better heading performance over the adaptive backstepping controller. For the towing condition, the non-adaptive backstepping controller shows 15% better heading performance over the adaptive controller.

The main findings of the set point, variable mass, variable drag, and variable mass and drag tests are that the surge speed performance of the adaptive controller is superior to the non-adaptive backstepping controller in the transitioning region. Finally, the adaptive controller shows less effects to changing mass and drag parameters than the backstepping controller, which is the ultimate goal of the low-level control for the ALR mission.

## VIII. CONCLUDING REMARKS

In this paper, we present the results of experimental testing of an unmanned surface vehicle under closed loop speed and heading control through lightship, transitional, and full displacement conditions. The tests were performed to evaluate the tracking performance of backstepping and model reference adaptive controllers under the variable mass and drag conditions expected during the automatic launch and recovery of an AUV from the USV. System identification of the data recorded from the water-borne maneuvering trials was used to find approximate values of hydrodynamic coefficients for the vehicle that adequately reproduce its motion in zig-zag and acceleration tests.

TABLE VII

COMPARISON OF THE ADAPTIVE BACKSTEPPING AND BACKSTEPPING CONTROLLERS FOR THE VARIABLE MASS AND DRAG TEST. USING THE METRIC IDENTIFIED IN SECTION VI.C, THE STEADY STATE SPEED ERROR OF THE ABS CONTROLLER IS 65% BETTER THAN THE BS CONTROLLER AFTER THE DROP. HOWEVER, THE STEADY STATE HEADING ERROR OF THE BS CONTROLLER IS 15% BETTER THAN THE ABS CONTROLLER AFTER THE DROP.

| Controller: | ABS | BS |
| --- | --- | --- |
| Steady State Speed Error Before Drop (m/s): | 0.071 | 0.14 |
| Percent Error: | 7% | 14% |
| Steady State Speed Error After Drop (m/s): | 0.066 | 0.19 |
| Percent Error: | 6% | 19% |
| Steady State Heading Error Before Drop (deg): | 3.1 | 0.36 |
| Steady State Heading Error After Drop (deg): | 3.4 | 2.9 |

We modified the backstepping controller for mono-hull vessels developed by Liao et al. (2010) [24] to account for the twin hull dynamics and actuator saturation limits of a wave-adaptive modular vessel. The modified controller was then used as the basis for a direct method model reference adaptive controller [17]. The reference model was derived using a physics-based model of the vessel that was tuned using system identification test data. We performed a comparative analysis of the steady state tracking performance between the two controllers (TABLE VII). Results show that the adaptive controller is superior in surge speed tracking performance – the improvement comes with a slight penalty in the effectiveness of heading control. The variable displacement and drag experiments show that the vehicle exhibits steady state surge tracking errors when an adaptive control approach is not used. The variable mass and drag experiment is the most representative of an ALR mission, and clearly demonstrates that the adaptive backstepping controller outperformed the non-adaptive backstepping controller in terms of speed and exhibited a similar response as the non-adaptive backstepping case in heading. These same results are illustrative of the separate variable mass and variable drag test, where the adaptive backstepping controller clearly outperformed in speed, while displaying a common performance in heading.

REFERENCES

[1] Aguiar, A. P. and Hespanha, J. P. "Logic-based switching control for trajectory-tracking and path-following of underactuated autonomous vehicles with parametric modeling uncertainty." In *Proc. of the 2004American Control Conference.* Vol. 4, pp. 3004-3010, Boston, MA, USA, 2004.

[2] Aguiar, A. P. and Pascoal, A. M. Dynamic positioning and way-point tracking of underactuated AUVs in the presence of ocean currents. *International Journal of Control* vol. 80, no. 7, pp.1092-1108, 2007.

[3] Alvarez, J., Bertaska, I., and von Ellenrieder, K. "Nonlinear Control of an Amphibious Vehicle." Proc. ASME *Dynamic Systems Control Conference,* Stanford, CA USA, 2013.

[4] Annamalai, A. S. K., Sutton, R., Yang, C., Culverhouse, P. and Sharma, S. Robust Adaptive Control of an Uninhabited Surface Vehicle. *J. Intelligent & Robotic Systems*, May 2014. DOI 10.1007/s10846-014-0057-2.

[5] Ashrafiuon, H. Muske, K. R. and McNinch, L.C., "Review of Nonlinear Tracking and Setpoint Control Approaches for Autonmous Underactuated Marine Vehicles," in *American Control Conference*, Baltimore, MD. USA, 2010.

[6] Ashrafiuoun H. Muske, K. R. McNinch, L. C. and Soltan, R. A. Sliding-mode Tracking Control of Surface Vessels. *IEEE Transactions of Industrial Electronics*, vol. 55, no. 11, pp. 4004-4011, 2008.

[7]    Bertaska, I., Alvarez, J., et al., "Experimental Evaluation of Approach Behavior for Autonomous Surface Vehicles." Proc. ASME *Dynamic Systems Control Conference,* Stanford, CA USA, 2013.

[8]    Bertaska, I.R. and von Ellenrieder, K.D., " Supervisory Switching Control of an Unmanned Surface Vehicle," *Proc. MTS/IEEE OCEANS'15* Washington DC, USA pp.1-10, 2015.

[9]    Bertaska, I.R, et al. "Experimental Evaluation of Automatically Generated Behaviors for USV Operations," *J. Ocean Engineering¸* vol. 106, pp. 496-514, 2015.

[10]  Caccia, M. Bibuli, M. Bono, R. and Bruzzone, G. Basic navigation, guidance and control of an unmanned surface vehicle. *Autonomous Robots*, vol. 25, no. 4, pp. 349-365, 2008.

[11]  Dhanak, M. R., Ananthakrishnan, P., Frankenfield, J., & von Ellenrieder, K. "Seakeeping Characteristics of a Wave-Adaptive Modular Unmanned Surface Vehicle." *ASME 2013 32nd Intl Conf on Ocean, Offshore and Arctic Engineering*. American Society of Mechanical Engineers, 2013.

[12]  Duerr, P. and von Ellenrieder, K.D., Scaling and Numerical Analysis of Nonuniform Waterjet Pump Inflows, *IEEE J. Oceanic Engineering,* vol.40, no.3, pp.701-709, 2015.

[13]  Faltinsen, O. M. *Hydrodynamics of high-speed marine vehicles*. Cambridge University Press, 2005.

[14]  Fossen, T. I. *Guidance and Control of Ocean Vehicles*. Chichester: John Wiley and Sons Ltd, 1994.

[15]  Fossen, T. I. and Strand, J. P. Tutorial on nonlinear backstepping: applications to ship control. *Modeling, Identification and Control*, vol. 20, no. 2, pp. 83-134, 1999.

[16]  Fratello, J. and Ahmadian, M. "Multi-body Dynamic Simulation and Analysis of Wave-adaptive Modular Vessels." *Proc. 11th Intl Conf on Fast Sea Transportation FAST 2011*, Honolulu, Hawaii, USA, 2011.

[17]  Ghommam, J. Mnif, F. Benali, A. and Derbel, N. Asymptotic backstepping stabilization of an underactuated surface vessel. *Control Systems Technology, IEEE Transactions on*, vol. 14, no. 3, pp. 1150-1157, 2006.

[18]  International Maritime Organization. "Explanatory Notes to the Standards for Ship Manoeuvrability." December 2002. London.

[19]  Khalil, H.K. *Nonlinear Systems*, Upper Saddle River, NJ: Prentice Hall, 2002.

[20]  Klinger, W. Bertaska, I. Alvarez, J. and von Ellenrieder, Karl, "Controller Design Challenges for Waterjet Propelled Unmanned Surface Vehicles with Uncertain Drag and Mass Properties." *Proc. MTS/IEEE Oceans'13* San Diego, CA. pp. 23-26. 2013.

[21]  Klinger, W. Bertaska and von Ellenrieder, K. D. "Experimental Testing of an Adaptive Controller for USVs with Uncertain Displacement and Drag." *Proc. MTS/IEEE Oceans'14* St. John's, NS Canada. 2014.

[22]  Kragelund, S, et al. "Adaptive Speed Control for Autonomous Surface Vessels," in *IEEE/MTS Oceans Conference 2013 (Oceans '13)*, San Diego, CA, USA. 2013.

[23]  Li. J, et al. "Robust Adaptive Backstepping Design for Course-keeping Control of Ship with Parameter Uncertainty and Input Saturation," in *2011 International Conference on Soft Computing and Pattern Recognition (SoCPaR)*, Dalian, Chiana, 2011.

[24]  Liao, Y. Pang, Y. and Wan, L. "Combined speed and yaw control of underactuated unmanned surface vehicles," *2010 2nd International Asia Conference on  Informatics in Control, Automation and Robotics (CAR)*, vol.1, pp. 157-161, March 2010.

[25]  Mahini, F. and Ashrafiuon H, "Autonomous Surface Vessel Target Tracking Experiments in Simulated Rough Sea Conditions," in *ASME Dynamci Systems and Control Conference*, Ft. Lauderdale, FL, USA 2012.

[26]  Marquardt, J. G. Alvarez, J. and von Ellenrieder, K. D. (2014) Characterization and System Identification of an Unmanned Amphibious Tracked Vehicle. *IEEE J Oceanic Engineering,* **39**(4):641-661.

[27]  Miranda, M., Beaujean, P. P., An, E., & Dhanak, M.  "Homing an Unmanned Underwater Vehicle Equipped with a DUSBL to an Unmanned Surface Platform: A Feasibility Study." *Proc. MTS/IEEE Oceans'13* San Diego, CA. pp. 23-26. 2013.

[28]  Panteley, E. and Loria, A. Growth rate conditions for uniform asymptotic stability of cascaded time-varying systems. *Automatica*, vol. 37, no. 3, 453-460, 2001.

[29]  Pearson, D. An, P.-C., Dhanak, M.R von Ellenrieder, K. D. and Beaujean, P.-P. "High-level fuzzy logic guidance system for an unmanned surface vehicle (USV) tasked to perform autonomous launch and recovery (ALR) of an autonomous underwater vehicle (AUV)." *Proc. 2014 IEEE/OES Autonomous Underwater Vehicles (AUV)* Oxford, MS pp.1-15, 2014.

[30]  Peterson, A. Ahmadian, M. Craft, M. and Shen, A. "Simulation and scale-testing to improve the next generation of wave-adaptive modular vessels." *Proc. 2013 Grand Challenges on Modeling and Simulation Conference*, p. 19. Society for Modeling & Simulation International, 2013.

[31]  Qu, H. Sarda, E.I. Bertaska, I.R. and von Ellenrieder, K.D., "Wind feed-forward control of a USV," *Proc. MTS/IEEE OCEANS'15* Genova, Italy pp.1-10, 2015.

[32]  Qu, H. Sarda, E.I. Bertaska, I.R. and von Ellenrieder, K.D., "Adaptive Wind Feedforward Control of an Unmanned Surface Vehicle for Station Keeping," *Proc. MTS/IEEE OCEANS'15* Washington DC, USA pp.1-10, 2015.

[33]  Sarda, E. and Dhanak, M. R. "Unmanned recovery of an AUV from a surface platform." *Proc. MTS/IEEE Oceans'13* San Diego, CA. pp. 23-26. 2013.

[34]  Sarda, E.I. Dhanak, M. R. and von Ellenrieder, K. D. "Concept for a USV–based autonomous launch and recovery system," *Proc. ASNE Launch & Recovery Symp. 2014*, Linthicum, MD USA, November 2014.

[35]  Sarda, E.I. Bertaska, I.R. Qu, A. and von Ellenrieder, K.D., "Development of a USV station-keeping controller," *Proc. MTS/IEEE OCEANS'15* Genova, Italy pp.1-10, 2015.

[36] Skjetne, R., Smogeli, O.N., and Fossen T.I. "A Nonlinear Ship Manoeuvering Model: Identification and Adaptive Control With Experiments for a Model Ship," *Modeling, Identification, and Control*, vol 25, no. 1, pp 3-27, 2004.

[37] Slotine, J.-J. E. and Li, W.-P. *Applied Nonlinear Control.* New Jersey: Prentice Hall, 1991.

[38] SNAME "Nomenclature for Treating the Motion of a Submerged Body through a Fluid". *The Society of Naval Architects and Marine Engineers, Technical and Research Bulletin No. 1-5*, pp. 1-15. April 1950.

[39] Sonnenburg, C.R., and Woolsey, A.C. "Modeling, Identification, and Control of an Unmanned Surface Vehicle," *J. Field Robotics*, vol 30, no. 3, pp1-28, 2013.

[40] Svec, Peter, Brual C. Shah, Ivan R. Bertaska, Jose Alvarez, Armando J. Sinisterra, Karl von Ellenrieder, Manhar Dhanak, and S. K. Gupta. "Dynamics-aware target following for an autonomous surface vehicle operating under COLREGs in civilian traffic." In *Intelligent Robots and Systems (IROS), 2013 IEEE/RSJ International Conference on*, pp. 3871-3878. IEEE, 2013.

[41] von Ellenrieder, K. D. Free Running Tests of a Waterjet Propelled Unmanned Surface Vehicle. *J. Mar. Engng & Tech* **12**(1):1-9, 2013.

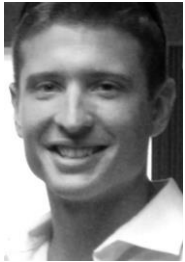

Wilhelm Klinger received the B.S. degree in ocean engineering from the U.S. Naval Academy, Annapolis, MD USA in 2012 and M.S. degree in ocean engineering from Florida Atlantic University, Boca Raton, FL USA in 2014. He worked as a Research Assistant at Florida Atlantic University's Seatech Institute for Ocean and Systems Engineering, Dania Beach, FL USA while completing his M.S. degree and is currently in flight training for the US Navy. Mr. Klinger is a member of the Marine Technology Society and Tau Beta Pi Engineering Honor Society.

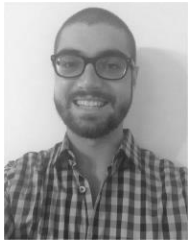

Ivan Bertaska (M'13) received the B.S. and M.S. degree in ocean engineering from Florida Atlantic University, Boca Raton, FL USA in 2012 and 2013, respectively. He is currently a Ph.D. candidate at Florida Atlantic University's Seatech Institute for Ocean and Systems Engineering, Dania Beach, FL USA. Mr. Bertaska is a member of the Tau Beta Pi Engineering Honor Society and the IEEE Oceanic Engineering Society. His areas of interest include unmanned vehicle systems control, software architecture and navigation.

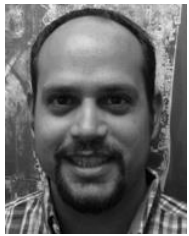

Karl von Ellenrieder (M'08) received the B.S. degree in aeronautics and astronautics from the Massachusetts Institute of Technology (MIT), Cambridge, MA USA in 1990 and the M.S. and Ph.D. degrees in aeronautics and astronautics from Stanford University, Stanford, CA USA in 1992 and 1998, respectively. He is currently Associate Director of The Institute for Ocean and Systems Engineering (SeaTech),and a Professor in the Department of Ocean & Mechanical Engineering at Florida Atlantic University in Dania Beach, FL. His research interests include unmanned surface vehicles, marine hydrodynamics and ship design. Prof. von Ellenrieder is a Member of the Society of Naval Architects and Marine Engineers, the Marine Technology Society and the IEEE Oceanic Engineering Society.

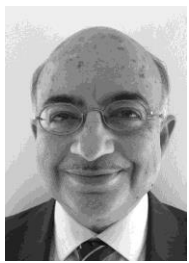

Manhar Dhanak is the Director of The Institute for Ocean and Systems Engineering (SeaTech), Director of Research for Southeast National Marine Renewable Energy Center and Professor of Ocean Engineering at Florida Atlantic University. He is the past chair (2003-2009) of the Department of Ocean Engineering at FAU. He is a graduate of Imperial College, University of London and served as a Research Scientist at Topexpress Ltd., Cambridge, UK and as a Senior Research Associate at University of Cambridge before joining FAU. Dr. Dhanak has research interests in hydrodynamics, physical oceanography, autonomous underwater vehicles (AUV) and ocean energy.